\documentclass[pra, onecolumn,10pt,aps,superscriptaddress,longbibliography]{revtex4-2}

\usepackage{graphicx}
\usepackage{dcolumn}
\usepackage{bm}
\usepackage{amssymb}
\usepackage{amsmath}
\usepackage{mathtools}
\usepackage{physics}
\usepackage[usenames,dvipsnames]{color}
\usepackage{hyperref}
\usepackage[normalem]{ulem}
\usepackage{microtype}

\newcommand{\rev}[1]{{\color{black}#1}}
\newcommand{\pro}[1]{{\color{black}#1}}
\newcommand{\kq}[1]{{\color{black}#1}}

\begin{document}

\title{\pro{Single-particle limit of a topological edge state 
in a locally resonant band gap}}

\author{Garigipati Sai Srikanth}
\thanks{These authors contributed equally.}
\affiliation{Department of Aerospace Engineering, Indian Institute of Science, Bangalore 560012, India}

\author{Kai Qian}
\thanks{These authors contributed equally.}
\affiliation{Department of Mechanical and Aerospace Engineering, University of California, San Diego, La Jolla, CA 92093, USA}
\affiliation{George W. Woodruff School of Mechanical Engineering, Georgia Institute of Technology, Atlanta, GA 30332, USA}

\author{Ian Frankel}
\thanks{These authors contributed equally.}
\affiliation{Department of Mechanical and Aerospace Engineering, University of California, San Diego, La Jolla, CA 92093, USA}

\author{Georgios Theocharis}
\affiliation{Laboratoire d’Acoustique de l’Université du Mans (LAUM), UMR 6613, Institut d’Acoustique -- Graduate School
(IA-GS), CNRS, Le Mans, France}

\author{Nicholas Boechler}
\affiliation{Department of Mechanical and Aerospace Engineering, University of California, San Diego, La Jolla, CA 92093, USA}
\affiliation{Program in Materials Science and Engineering, University of California, San Diego, La Jolla, CA 92093, USA}

\author{Rajesh Chaunsali}
\email{rchaunsali@iisc.ac.in}
\affiliation{Department of Aerospace Engineering, Indian Institute of Science, Bangalore 560012, India}

\date{\today}

\begin{abstract}
Topological metamaterials promise unprecedented wave control. Here, we theoretically and numerically investigate a one-dimensional Su-Schrieffer-Heeger (SSH)-inspired stiffness dimer modified with a local resonator, which imparts a frequency-dependent effective stiffness to the unit cell.
\pro{The resonator introduces an attenuation singularity: at a frequency at which the effective stiffness vanishes, the spatial attenuation of waves diverges. By tuning a dimerization parameter, we migrate this singularity from one band gap to the other via an intermediate flat-band state, transferring the dominant local-resonance character between the gaps without closing either gap and while preserving the underlying band topology.}
Crucially, \rev{when} the resulting topological edge state intersects \rev{the attenuation singularity, the edge state collapses onto a single boundary particle, forming a single-particle mode (SPM). This yields} an inverse participation ratio of exactly unity, the theoretical limit for localization in a discrete system. \pro{Moreover, this extreme localization can be realized at low frequencies, below the first Bragg-type band gap.}
Further, we demonstrate that while random disorder \rev{detunes} this mode, \rev{merely tuning the} boundaries stabilizes the single-particle mode over a broad parameter range. Our findings provide a clear pathway to designing ultra-localized \rev{edge} states in low-frequency regimes\rev{, where band topology guarantees the edge mode and local resonance drives its single-particle confinement}.
\end{abstract}

\maketitle

\section{Introduction}

Engineered materials, such as phononic crystals and acoustic metamaterials, provide powerful platforms for controlling wave propagation. One of their defining features is the ability to create frequency band gaps---spectral ranges where wave propagation is attenuated. These band gaps typically arise through two distinct physical mechanisms. The first is Bragg scattering, where coherent scattering from a periodic lattice opens so-called Bragg-type band gaps (BrGs)~\cite{lu2009phononic}. The second is local resonance, achieved by embedding resonant substructures into a host medium. These local resonators give rise to effective material properties (\textit{e.g.}, dynamic mass density or modulus) that can vanish or diverge at characteristic frequencies, creating \rev{locally resonant} band gaps (LRGs)~\cite{li2004double,fang2006ultrasonic,chan2006extending,ding2007metamaterial,lee2016origin,wang2014dynamic}. \pro{Beyond their differing physical origins, the two gap types differ in two key respects.} \rev{Locally resonant band gaps can exhibit theoretically infinite attenuation (in contrast to the finite attenuation of BrGs for non-vanishing stiffnesses)~\cite{liu2012wave,theocharis2014limits,lin2021numerical}}. \pro{Locally resonant gaps can also be positioned at lower frequencies than BrGs while maintaining other properties, such as a high long-wavelength sound speed (wherein the resonators can be deeply ``sub-wavelength''~\cite{achaoui2011experimental,jamil2022inerter} and the gap frequency ``sub-Bragg''~\cite{raghavan2013local}).}

Parallel to these developments, the introduction of topological band theory into mechanical systems has catalyzed the vibrant field of topological mechanics~\cite{liu2020topological,Colloquium2024}. In these systems, robust boundary modes (or states) emerge at edges~\cite{Wang2015}, interfaces~\cite{Mousavi2015}, and corners~\cite{Serra-Garcia2018}. Residing deep within band gaps, these states exhibit remarkable resilience to disorder and imperfections due to their underlying topological protection. This intrinsic robustness makes them prime candidates for next-generation applications in fault-tolerant, one-way passive waveguiding, logic, and signal processing. \rev{These topological boundary modes pair naturally with local resonance; the sharp attenuation of LRGs offers a route to edge localization complementary to Bragg-gap confinement. While BrG edge modes decay exponentially over broad parameter ranges, an LRG introduces an attenuation singularity that can drive extreme confinement at a specific frequency, motivating our study of resonance-assisted edge localization in topological metamaterials.}

\begin{figure}[!]
\includegraphics[width=0.9\textwidth]{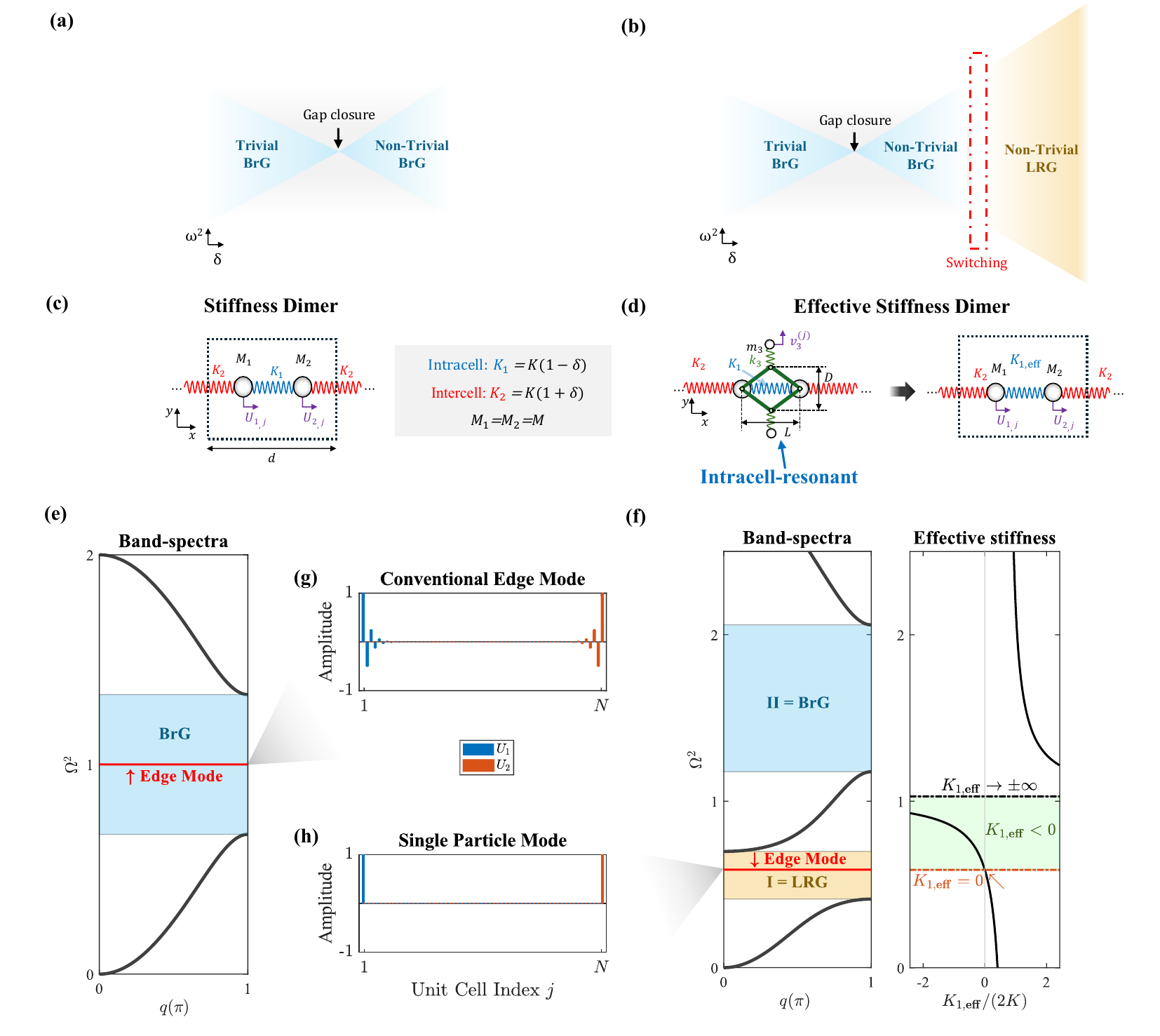}
\caption{\textbf{Comparison of topological edge modes in a Bragg-type band gap (BrG) and an effective stiffness\rev{, locally resonant} band gap (LRG).}
(a) Band closure and reopening driving topological transition in a BrG.
(b) Pathway to topologically non-trivial LRG via gap-type switching from a non-trivial BrG.
(c) Unit cell of conventional stiffness dimer chain.
(d) Unit cell of effective stiffness dimer chain with local resonators ($m_3, k_3$).
(e) Band structure of conventional dimer; blue region denotes BrG, red line indicates edge mode.
(f) Left: Band structure of effective stiffness dimer with BrG (blue, II) and LRG (yellow, I). Right: $K_{1,\text{eff}}$ versus {a normalized $\Omega^2=\omega^2M/(K_1+K_2)$}; orange dashed line marks $K_{1,\text{eff}}=0$.
(g) Edge mode profile in conventional dimer showing exponential decay.
(h) Single-particle mode (SPM), localized on the edge with IPR$=1$, in the effective stiffness dimer.
Parameters: ${K_1=1}$, ${K_2=1.4286}$, ${M=1}$, ${k_3=0.5}$, ${m_3=0.2}$, ${\lambda=1}$, ${N=60}$ unit cells with fixed boundaries.}
\label{fig:intro}
\end{figure}

Localized boundary modes \rev{with topological origin} have been previously reported in \rev{band gaps stemming from local resonance}. For example, Zhang \textit{et al.}~\cite{zhang2020dirac} introduced resonator pairs into a classical valley-Hall lattice to obtain edge modes in resonant ranges, and Xia \textit{et al.}~\cite{xia2020topological} demonstrated edge modes in a beam with quasi-periodically attached resonators. \rev{However, until recently, the mechanism governing the topological transition from a trivial to a non-trivial LRG remained unclear.} Conventional topological states at finite frequencies arise through band inversion at Dirac points, requiring gap closure and reopening [Fig.~\ref{fig:intro}(a)]. \pro{In LRGs, by contrast,} the attenuation singularity---associated with vanishing effective stiffness (equivalently, diverging effective mass)---enforces a non-vanishing minimum gap width~\cite{sugino2016mechanism}, thereby preventing gap closure under continuous parameter tuning~\cite{zhao2018topological}.
A recent study by Jang \textit{et al.}~\cite{JANG2025} resolved this through a two-step process: first inducing a topological transition within a BrG, and subsequently switching its gap character to an LRG, thus bypassing the need for gap closure within the LRG itself [Fig.~\ref{fig:intro}(b)]. Similarly, Zhang \textit{et al.}~\cite{zhang2024realization} illustrated a BrG-to-LRG transition in a beam-resonator system. \rev{Notably, Jang \textit{et al.}~\cite{JANG2025} also showed that the topological edge state formed at the local-resonance frequency, where the effective mass diverges, is confined to a single unit cell, forming a \textit{singular edge mode}. This raises the questions that we address here: what is the theoretical limit of such localization---can a topological edge state be confined to a single \textit{particle} (\textit{i.e.}, sub-unit-cell localization)?} \pro{Further, can this be achieved in a low-frequency LRG, below the first BrG?}

\rev{We answer these questions with} a three-band effective stiffness model\rev{---namely,} a modified version of the conventional SSH-inspired stiffness dimer chain [\rev{the main chain, }composed of equal masses connected by alternating springs, as shown in Fig.~\ref{fig:intro}(c)], where a truss-based local resonator is attached to the intracell spring [as shown in Fig.~\ref{fig:intro}(d), which we refer to as the \textit{effective stiffness dimer chain}].
\pro{We show that, as a stiffness-dimerization parameter is swept, the attenuation singularity (zero effective stiffness) introduced by the resonator crosses from one band gap to the other precisely when an intervening flat band forms, transferring the dominant local-resonance character between the gaps while preserving the band topology. The resulting topologically non-trivial gap, which we will herein refer to as an LRG (see Sec.~II\,B for the subtleties of this terminology), hosts topological edge modes.} \rev{When such an} edge state intersects \rev{the attenuation} singularity, it becomes confined to a single particle at the boundary, \rev{forming what we term the \textit{single-particle mode} (SPM)---\pro{the sub-unit-cell limiting case} of the singular edge mode}. This \rev{confinement} represents the theoretical maximum of localization in a discrete system and, to the best of our knowledge, has not been demonstrated \rev{previously}. \rev{Further, we are able to achieve this SPM at low frequencies, \pro{below the first BrG,} by leveraging the aforementioned advantages conferred by \pro{local resonance}.} \rev{Finally, we demonstrate that, by simple tuning of the boundary stiffness, the SPM can be maintained over a finite range of unit-cell parameters, rather than only at a fine-tuned resonance condition, even with random disorder in the chain.}

The remainder of this paper is organized as follows. Section~II presents the results: Sec.~II\,A introduces the unit-cell dynamics and effective model; Sec.~II\,B details band gap characterization and the switching mechanism; Sec.~II\,C analyzes the topological transition via band inversion; Sec.~II\,D examines the finite chain spectrum and edge state localization; and Sec.~II\,E discusses disorder analysis and tuned boundaries. Section~III concludes with a summary and outlook.

\section{Results}

\subsection{Unit-cell dynamics and effective model}

We begin by introducing the model, its equations of motion, and the corresponding dynamical matrix. Our system is a quasi-one-dimensional diatomic lattice, as illustrated in Fig.~\ref{fig:intro}(d). \rev{The $j$th} unit cell consists of two identical primary masses $M$, constrained to horizontal motion, with displacements denoted by \rev{$U_{1,j}$} and \rev{$U_{2,j}$}. The primary masses are coupled by alternating linear springs of stiffness $K_1$ (intracell) and $K_2$ (intercell). The intracell spring is modified by a symmetric truss-type local-resonator structure proposed in Ref.~\cite{huang2011theoretical}. This structure comprises a pair of identical local resonators, each consisting of a mass $m_3$ and a spring $k_3$, coupled to the main chain through a hinged, massless truss mechanism. The resonator masses are constrained to vertical motion with displacement denoted by \rev{$v_{3,j}$}. As a result, the intracell coupling becomes frequency dependent.

The linearized equations of motion governing the $j$th unit cell are given by:
\begin{align}
    \label{eq:eom1}
    M \ddot{U}_{1,j} + K_1(U_{1,j}-U_{2,j}) + K_2(U_{1,j}-U_{2,j-1})-k_3\lambda\left[v_{3,j}+\frac{\lambda}{2}(U_{2,j}-U_{1,j})\right] &= 0, \\
    \label{eq:eom2}
    M \ddot{U}_{2,j} + K_1(U_{2,j}-U_{1,j}) + K_2(U_{2,j}-U_{1,j+1})+ k_3\lambda\left[v_{3,j}+\frac{\lambda}{2}(U_{2,j}-U_{1,j})\right] &= 0, \\
    \label{eq:eom3}
    m_3 \ddot{v}_{3,j} + k_3\left[v_{3,j}+\frac{\lambda}{2}(U_{2,j}-U_{1,j})\right] &= 0,
\end{align}
where $\lambda=L/D$ characterizes the geometry of the truss mechanism, relating the horizontal displacement of the primary masses to the vertical displacement of the resonator (see Sec.~I\,A of the Supplemental Material for geometric details). The terms involving $k_3\lambda$ represent the coupling forces transmitted through the truss mechanism.

Applying Bloch's theorem with the ansatz \rev{$[U_{1,j}, U_{2,j}, v_{3,j}]^T = [U_1, U_2, v_3]^T\, e^{i(qj - \omega t)}$} transforms these equations into a $3\times3$ eigenvalue problem, $\mathbf{D}(q)\mathbf{U} = \omega^2 \mathbf{U}$, where $\mathbf{U} = [U_1, U_2, \sqrt{2m_3/M}\,v_3]^T$ is the eigenvector of modal amplitudes. \rev{Here, $U_1$, $U_2$, and $v_3$ denote the Bloch amplitudes corresponding to $U_{1,j}$, $U_{2,j}$, and $v_{3,j}$, respectively; the factor $\sqrt{2m_3/M}$ is included to make $\mathbf{D}(q)$ Hermitian.}
The full dynamical matrix $\mathbf{D}(q)$ is:
\begin{equation}
    \label{eq:Dmatrix_intracell}
    \mathbf{D}(q) =
    \begin{pmatrix}
    \dfrac{K_1+K_2}{M}+\dfrac{k_3\lambda^2}{2M}
    &
    -\left(\dfrac{K_1}{M}+\dfrac{K_2}{M}e^{-iq}+\dfrac{k_3\lambda^2}{2M}\right)
    &
    -\dfrac{k_3 \lambda}{\sqrt{2Mm_3}} \\[2ex]
    -\left(\dfrac{K_1}{M}+\dfrac{K_2}{M}e^{iq}+\dfrac{k_3\lambda^2}{2M}\right)
    &
    \dfrac{K_1+K_2}{M}+\dfrac{k_3\lambda^2}{2M}
    &
    \dfrac{k_3 \lambda}{\sqrt{2Mm_3}} \\[2ex]
    -\dfrac{k_3\lambda}{\sqrt{2Mm_3}} & \dfrac{k_3\lambda}{\sqrt{2Mm_3}} & \dfrac{k_3}{m_3}
\end{pmatrix},
\end{equation}
\rev{where $q=\tilde{\nu}d$ is the normalized wavenumber, $d$ is the lattice constant, and $\tilde{\nu}$ is the physical wavenumber.}
The diagonal elements represent on-site stiffness contributions, while the off-diagonal elements encode coupling between degrees of freedom (DoF).

To facilitate comparison with a standard stiffness dimer and gain deeper physical insight, we employ an effective medium framework. The local resonator dynamics [Eq.~(\ref{eq:eom3})] can be solved algebraically for harmonic motion at frequency $\omega$:
\begin{equation}
    v_3 = -\frac{\lambda}{2}\frac{\omega_r^2}{\omega_r^2 - \omega^2}(U_2 - U_1),
    \label{eq:v3_solution}
\end{equation}
where $\omega_r = \sqrt{k_3/m_3}$ is the natural frequency of the local resonators. Substituting this back into Eqs.~(\ref{eq:eom1}) and (\ref{eq:eom2}), the entire intracell mechanical structure can be described by a single, frequency-dependent effective stiffness $K_{1,\text{eff}}(\omega)$. Following Ref.~\cite{huang2011theoretical}:
\begin{equation}
    \label{eq:K1eff}
    K_{1,\text{eff}}(\omega)
    = K_1 + \frac{\lambda^2 k_3}{2}\frac{\omega^2}{\omega^2-\omega_r^2}.
\end{equation}
Equation~\eqref{eq:K1eff} captures the essential physics: below resonance ($\omega < \omega_r$), the second term is negative and reduces the effective stiffness; above resonance ($\omega > \omega_r$), it is positive and increases the stiffness. At $\omega = \omega_r$, the effective stiffness diverges ($K_{1,\text{eff}} \to \pm\infty$)\pro{; it vanishes ($K_{1,\mathrm{eff}}=0$) at a distinct frequency $\omega_s < \omega_r$, derived in the next section.}

By substituting Eq.~(\ref{eq:K1eff}), the 3-DoF system reduces to an effective 2-DoF model:
\begin{align}
    \label{eq:eomU1_K1eff}
    M \ddot{U}_{1,j} + K_{1,\text{eff}}(\omega)(U_{1,j}-U_{2,j}) + K_2(U_{1,j}-U_{2,j-1}) &= 0, \\
    \label{eq:eomU2_K1eff}
    M \ddot{U}_{2,j} + K_{1,\text{eff}}(\omega)(U_{2,j}-U_{1,j}) + K_2(U_{2,j}-U_{1,j+1}) &= 0.
\end{align}
Applying Bloch's theorem yields a $2\times2$ eigenvalue problem, $\mathbf{D}_{\text{eff}}(q, \omega)\mathbf{U}_{\text{eff}} = \omega^2 \mathbf{U}_{\text{eff}}$, where $\mathbf{U}_{\text{eff}} = [U_1, U_2]^T$ and:
\begin{equation}
    \label{eq:Dmatrix_intracell_2DoF}
    \mathbf{D}_{\text{eff}}(q, \omega) =
    \frac{1}{M}
    \begin{pmatrix}
    K_{1,\text{eff}}+K_2 & -(K_{1,\text{eff}}+K_2 e^{-iq}) \\[1ex]
    -(K_{1,\text{eff}}+K_2 e^{iq}) & K_{1,\text{eff}}+K_2
    \end{pmatrix}.
\end{equation}
This $2\times2$ formulation is structurally identical to the standard SSH model, with the crucial difference that $K_{1,\text{eff}}$ is frequency dependent. \rev{The band structure nonetheless remains well defined at fixed structural parameters; the full $3\times3$ Bloch \pro{dynamical matrix} [Eq.~\eqref{eq:Dmatrix_intracell}] contains no frequency-dependent parameters and yields \pro{these bands} directly as a standard linear eigenvalue problem. The frequency dependence of $K_{1,\text{eff}}$ arises only from eliminating the resonator coordinate; the resulting $2\times2$ form [Eq.~\eqref{eq:Dmatrix_intracell_2DoF}] is implicit in $\omega$ but, solved self-consistently, reproduces the same bands.}

\subsection{Band gap characterization and switching}

We now analyze the band structure to understand wave propagation \rev{in our system} and detail the mechanism for switching the band gap type [Fig.~\ref{fig:intro}(b)].
To facilitate a systematic parameter study, we introduce a dimerization parameter $\delta$, \rev{treated as a static structural parameter,} and a constant stiffness scale $K$ \rev{to define the physical alternating stiffnesses as}
\begin{equation}
    K_1 = K(1-\delta), \quad K_2 = K(1+\delta),
    \label{eq:K1K2_parametrization}
\end{equation}
with $K>0$ and $\delta\in(-1,1)$.
This isolates $\delta$ as the primary control parameter: positive $\delta$ corresponds to weaker intracell coupling relative to intercell coupling, and vice versa.
We further introduce non-dimensional quantities to express our results in universal form: stiffness ratio $\kappa_3 = k_3/(2K)$; characteristic frequency $\omega_{\text{mid}} = \sqrt{2K/M}$ (the mid-gap frequency of the unit cell without local resonators at $\delta = 0$); normalized resonator frequency $\Omega_r = \omega_r/\omega_{\text{mid}}$; and normalized frequency $\Omega = \omega/\omega_{\text{mid}}$, such that $\Omega^2 = \omega^2 M/(2K)$.

\begin{figure}[t!]
    \centering
    \includegraphics[width=0.8\textwidth]{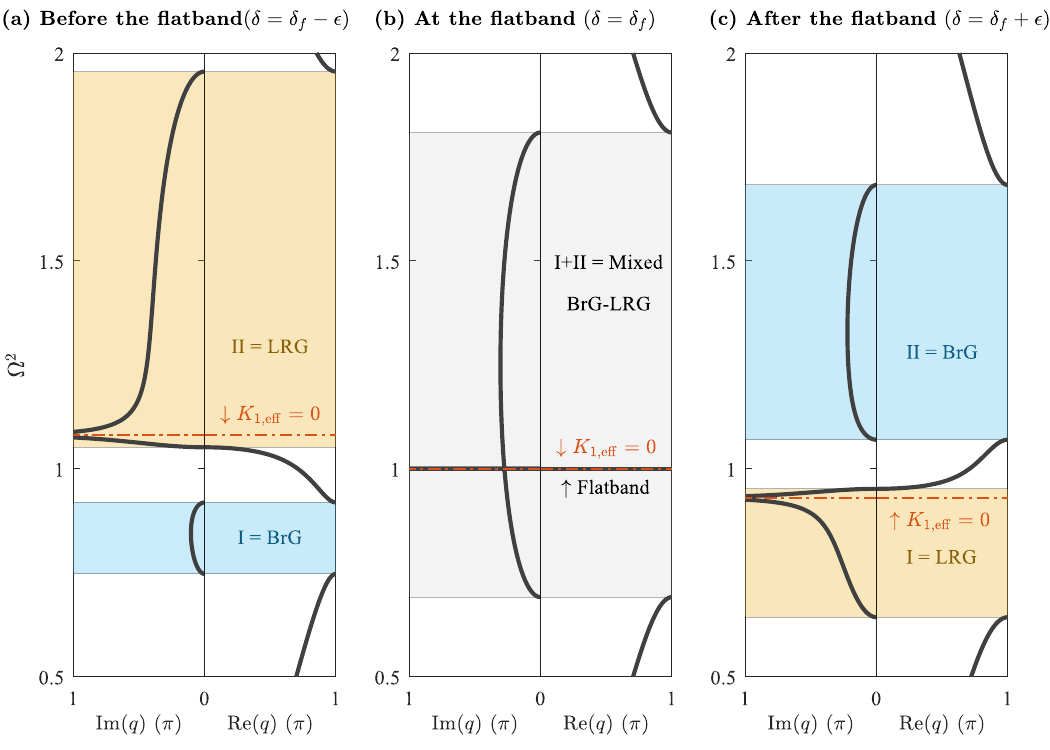}
 \caption{\textbf{Band gap type switching via a flat band.}
    Orange dashed line marks $K_{1,\text{eff}}=0$; its location defines LRG (yellow) versus BrG (blue).
    (a) $\delta < \delta_f$: singularity in upper gap (II = LRG).
    (b) $\delta = \delta_f$: singularity traversal through the flat band \pro{(the transitional `mixed' BrG--LRG state)}.
    (c) $\delta > \delta_f$: singularity in lower gap (I = LRG).
    Offset $\epsilon = 0.15\delta_f$ used in (a) and (c). Parameters: $K=1$, $M=1$, $k_3=0.5$, $m_3=0.2$, $\lambda=1$.
}
\label{fig:aroundflat}
\end{figure}

\pro{Before analyzing the switching mechanism, we clarify how we classify the band gaps. In their pure limits, the two gap types are cleanly separable: a pure BrG is set by the periodicity of the main chain and survives removal of the resonators, whereas a pure LRG is pinned to the resonator frequency and is insensitive to changes in the lattice constant~\cite{liu2012wave,kaina2013composite}. At finite coupling between the resonators and the main chain, however, both mechanisms contribute to every gap, and each gap generically acquires a hybrid character; prior works, such as Ref.~\cite{liu2012wave}, describe criteria for classifying such gaps. Here, we classify each gap by its dominant mechanism, using the attenuation singularity---a frequency at which the spatial decay rate of evanescent waves diverges ($\mathrm{Im}(q)\to\infty$)---as the discriminant: a gap containing the singularity is local-resonance dominant, and we call it an LRG; a gap lacking it is Bragg dominant, and we call it a BrG. 
We adopt these operational definitions throughout this study.} \rev{In our system, the} attenuation singularity corresponds precisely to the vanishing of the effective stiffness, $K_{1,\text{eff}} = 0$. To find this singular frequency, we set Eq.~(\ref{eq:K1eff}) to zero and solve for $\omega_s^2$:
\begin{equation}
    \omega_s^2 = \omega_r^2 \frac{K_1}{K_1 + \lambda^2 k_3/2} = \omega_r^2 \left(1+\frac{\lambda^2 k_3}{2K_1}\right)^{-1}.
\end{equation}
Substituting $K_1 = K(1-\delta)$ and $k_3 = 2K\kappa_3$, and normalizing by $\omega_{\text{mid}}^2 = 2K/M$, we obtain:
\begin{equation}
    \label{eq:omegaKeff=0}
    \Omega_s = \Omega_r \left(1+\frac{\lambda^2 \kappa_3}{1-\delta}\right)^{-1/2}.
\end{equation}
\rev{Here, $\Omega_s$ denotes the normalized zero-effective-stiffness frequency associated with $K_{1,\mathrm{eff}}=0$.}
Physically, when $K_{1,\text{eff}} = 0$, the intracell coupling vanishes, effectively decoupling adjacent unit cells and preventing wave propagation entirely. \pro{Consistent with this classification, the gap containing $\Omega_s$ is thus the LRG.}
\pro{A conventional alternative for locating the BrG is the Bragg periodicity frequency $\Omega_p$: the frequency at which a wave traveling at the long-wavelength sound speed of the chain has a wavelength equal to twice the lattice constant, so that Bragg scattering first becomes possible. We do not adopt this criterion because, once the two mechanisms overlap, $\Omega_p$ need not lie within the BrG (see Sec.~IV in the Supplemental Material, where $\Omega_p$ is derived and tested against $\Omega_s$).}

Figure~\ref{fig:aroundflat} tracks the complex band structure evolution as $\delta$ is varied. We first define a particular dimerization parameter,  $\delta_f$, where the middle branch forms a flat band~\cite{Leykam2018} with zero group velocity across the entire Brillouin zone. In Fig.~\ref{fig:aroundflat}(a), where $\delta < \delta_f$, the upper gap (II) contains the singularity (orange dashed line where $\text{Im}(q) \to \infty$), identifying it as an LRG. The lower gap (I) has a bounded, arch-shaped attenuation profile characteristic of Bragg scattering, identifying it as a BrG.
As $\delta$ increases, the singularity migrates from gap II to gap I\rev{, so the gap character switches in the sense that the local-resonance attenuation singularity is transferred between the two gaps}.
This interchange occurs at  $\delta = \delta_f$, derived by finding when the middle band becomes dispersionless (see Fig.~\ref{fig:aroundflat}(b) and the Supplemental Material):
\begin{equation}
\delta_{f} =  \left(\frac{\kappa_3\lambda^2+ \Omega_r^2}{2} \right) - \sqrt{\left(\frac{\kappa_3\lambda^2+ \Omega_r^2}{2} -1\right)^2 + 2 \lambda^2 \kappa_3}.
\label{eq:delta_f}
\end{equation}

This flat band is a necessary intermediate state for singularity migration as the stiffness parameter $\delta$ is swept. Within a propagating band, the wavenumber $q$ is strictly real, whereas the attenuation singularity belongs to the evanescent spectrum ($\text{Im}(q)\to\infty$)\rev{; a band of} finite bandwidth therefore acts as a spectral barrier that the singularity cannot traverse.
Only when the intervening band collapses to zero bandwidth---forming \rev{the flat band at $\delta=\delta_f$}---does this barrier vanish, allowing the singularity to cross from one gap to the other~\cite{banerjee2021inertial,frandsen2016inertial}.
\rev{At this switching point [Fig.~\ref{fig:aroundflat}(b)], gaps I and II are bridged only by the flat band and effectively merge into a single, larger gap, which we denote as a `mixed' BrG--LRG following the convention of Ref.~\cite{kaina2013composite} (as noted previously, the other gaps could also be considered `mixed' depending on the criteria used).}
For $\delta > \delta_f$ [Fig.~\ref{fig:aroundflat}(c)], the singularity then resides in gap I, making it the LRG, while gap II becomes a BrG\rev{---the reverse of the assignment for $\delta<\delta_f$}.
\rev{We emphasize that this flat-band--mediated migration is specific to the discrete model studied here; in continuum and flexural-beam systems, BrG--LRG switching has been reported without invoking flat-band formation~\cite{liu2012wave}.}

\pro{Finally, we note where these gaps lie relative to the Bragg scale. For
$\delta > \delta_f$, the singularity-hosting gap I sits below the BrG (gap II)
and, depending on the resonator parameters, can even lie entirely below
$\Omega_p$, \textit{i.e.}, below the frequency at which Bragg interference
first becomes possible; gaps in this regime are termed
``sub-Bragg''~\cite{raghavan2013local}. Since the long-wavelength sound speed
is set by the main chain alone (see Sec.~IV of the Supplemental Material), the
topological edge state analyzed below can reside in a low-frequency gap
obtained without sacrificing the quasi-static stiffness of the chain.}

\subsection{Topological transition via band inversion}

This section investigates the topological transitions arising from band gap closures---specifically, the formation of Dirac points---and the topological consequences of the gap-type switching discussed in the previous section. Our primary objective is to establish the conditions under which the LRG can host a topological phase.

To formally characterize the topology, we must first establish the relevant symmetries. The full $3\times 3$ dynamical matrix $\mathbf{D}(q)$ is inversion symmetric [satisfying $\mathbf{P}\mathbf{D}(q)\mathbf{P}^{-1}=\mathbf{D}(-q)$], with the parity operator given as {(see the Supplemental Material)}:
\begin{equation}
\label{eq:parity_operator}
 \mathbf{P} =
 \begin{pmatrix}
    0 & -1 & 0 \\
    -1 & 0 & 0 \\
    0 & 0 & 1
    \end{pmatrix}.
\end{equation}

\noindent This $3 \times 3$ structure complicates a direct mapping to the canonical Su-Schrieffer-Heeger (SSH) model. Here, the power of the effective medium framework becomes evident: the $2\times 2$ effective dynamical matrix $\mathbf{D}_{\text{eff}}(q)$ also preserves inversion symmetry, but in a form that allows us to utilize standard 1D topological invariants.

The inversion symmetry of the effective model is expressed as:
\begin{equation}
    \label{eq:inversion_2x2}
    \boldsymbol{\sigma}_x\mathbf{D}_{\text{eff}}(-q)\boldsymbol{\sigma}_x^{-1}=\mathbf{D}_{\text{eff}}(q),
\end{equation}
where the Pauli matrix $\boldsymbol{\sigma}_x$ acts as the inversion operator, swapping the two primary masses. At the high-symmetry points of the Brillouin zone ($q=0$ and $q=\pi$), time-reversal symmetry imposes $\mathbf{D}_{\text{eff}}(-q) = \mathbf{D}_{\text{eff}}^*(q) = \mathbf{D}_{\text{eff}}(q)$ since the matrix is real. Consequently, at these points, $\mathbf{D}_{\text{eff}}$ commutes with the inversion operator $\boldsymbol{\sigma}_x$. This commutation dictates that the eigenvectors of the \rev{dynamical matrix} \pro{can be chosen as} simultaneous eigenvectors of $\boldsymbol{\sigma}_x$. Since the eigenvalues of $\boldsymbol{\sigma}_x$ are $+1$ and $-1$, the eigenvectors at these high-symmetry points are constrained to be either symmetric (parity $+1$) or antisymmetric (parity $-1$):
\begin{equation}
    \mathbf{U}_{\text{eff}}^{\text{even}}  \propto \begin{pmatrix} 1 \\ 1 \end{pmatrix} \quad \text{or} \quad \mathbf{U}_{\text{eff}}^{\text{odd}} \propto \begin{pmatrix} 1 \\ -1 \end{pmatrix}.
\end{equation}

This symmetry constraint provides a direct physical observable to probe the topology. For a symmetric eigenvector, the two masses move in phase, and the phase difference is $|\arg(U_2)-\arg(U_1)| = 0$. For an antisymmetric eigenvector, they move in anti-phase, and the phase difference is $|\arg(U_2)-\arg(U_1)| = \pi$. Therefore, monitoring this phase difference across the Brillouin zone offers a direct map of the band's parity eigenvalues. A change in this quantized phase from $0$ to $\pi$ (or vice versa) between $q=0$ and $q=\pi$ signals a band inversion and a non-trivial Zak phase.

\begin{figure}[!]
    \centering
    \includegraphics[width=0.8\textwidth]{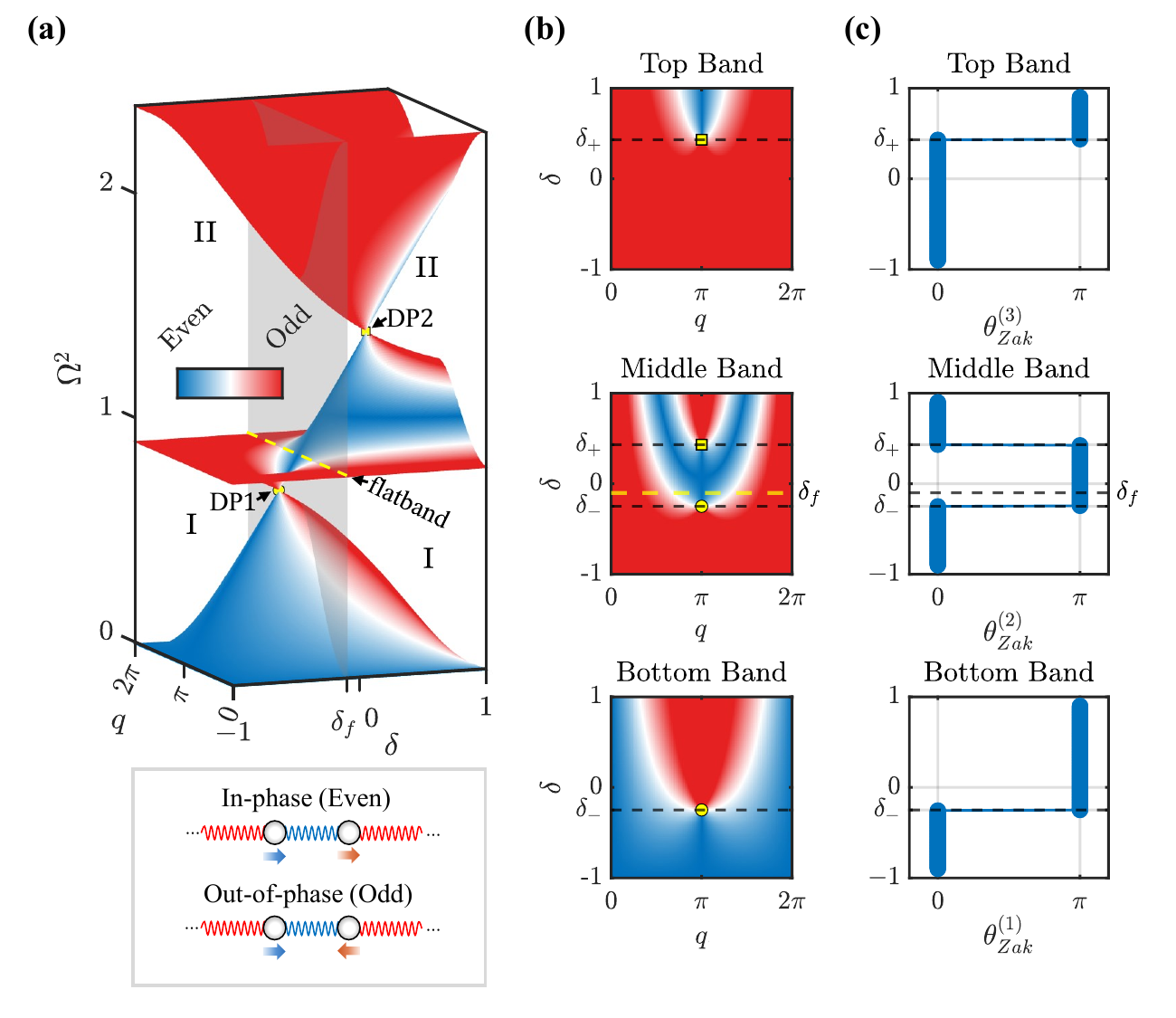}
 \caption{\textbf{Topological phase transitions and band inversion.}
    (a) Band structure evolution with $\delta$. Color map: phase difference $|\arg(U_2)-\arg(U_1)|$ {(in-phase = even = blue = 0, out-of-phase = odd = red = $\pi$)}. DP1/DP2 mark Dirac points; dashed yellow line and the vertical gray plane indicate the flatband and $\delta_f$, respectively.
    (b) Phase along the Brillouin zone versus $\delta$ for each band; phase change along $q=\pi$ shows inversion at $\delta_{\pm}$.
    (c) Zak phase versus $\delta$, confirming transitions only at $\delta_{\pm}$.
    Parameters: $K=1$, $M=1$, $k_3=0.43$, $m_3=0.2$, $\lambda=1$.
}
\label{fig:band_inversion}
\end{figure}

Figure~\ref{fig:band_inversion}(a) visualizes this topology by plotting the phase difference $|\arg(U_2)-\arg(U_1)|$ for all bands. We observe band gap closures at two distinct Dirac points located at $q=\pi$: The first Dirac point (DP1) occurs at $\delta = \delta_{-}$ (inverting the bottom and middle bands), and the second Dirac point (DP2) occurs at $\delta = \delta_{+}$ (inverting the middle and top bands). These critical values correspond to the condition where two bands touch each other at the Brillouin zone boundary (see the Supplemental Material for derivation):
\begin{equation}
    \delta_{\pm} = \left(\frac{\kappa_3\lambda^2}{4}+\frac{\Omega_r^2-1}{2}\right)\pm \sqrt{ \left(\frac{\kappa_3\lambda^2}{4}+\frac{\Omega_r^2-1}{2}\right)^2 + \frac{\lambda^2\kappa_3 }{2}}.
    \label{eq:delta_pm}
\end{equation}
The hierarchy of critical points is $\delta_{-} < \delta_{f} < \delta_{+}$. This establishes a clear sequence of transitions as $\delta$ increases: (i) a topological transition in the lower gap (I) via DP1 at $\delta_{-}$; (ii) the BrG-to-LRG character switch at $\delta_{f}$; and (iii) a topological transition in the upper gap (II) via DP2 at $\delta_{+}$.
{In Figure~\ref{fig:band_inversion}(b), phase change is seen, via a change in the color plotted, along $q=\pi$ at $\delta=\delta_-$ and $\delta=\delta_+$.}

To formally confirm these observations, we calculate the Zak phase for each band using the discrete formulation~\cite{vanderbilt2018berry}:
\begin{equation}
    \theta_{\text{Zak}}^{(b)} = -\text{Im}\ln{\left[\prod_{r=0}^{R-1} \langle\psi_{b}(q_{r})\mid\psi_{b}(q_{r+1})\rangle\right]}, \quad (\ket{\psi_{b}(q_{R})} \equiv \ket{\psi_{b}(q_{0})}),
    \label{eq:zak_phase}
\end{equation}
where $\ket{\psi_{b}}$ is the normalized eigenvector. As shown in Fig.~\ref{fig:band_inversion}(c), the Zak phase jumps between $0$ and $\pi$ exactly at $\delta_{\pm}$, matching the phase difference analysis. Finally, we define the gap topological invariant $\varsigma^{(b)}$ as the sum of Zak phases for all bands below the gap~\cite{xiao2014surface,zhao2018topological}:
\begin{equation}
    \text{sgn}[\varsigma^{(b)}]=(-1)^{b}\exp\left(i\sum_{m=1}^{b}\theta_{\text{Zak}}^{(m)}\right).
    \label{eq:gap_invariant}
\end{equation}
Table~\ref{table:intra} summarizes these invariants. The most crucial finding here is the topological inheritance: the gap-type switching at $\delta_f$ does \textit{not} alter the topological invariant. A gap that is topologically non-trivial as a BrG retains its non-trivial character even after it switches to become an LRG.

\begin{table}[h]
\centering
\caption{Sign of topological gap invariant $\varsigma^{(b)}$ for intracell-resonant stiffness dimer chains.}
\label{table:intra}
\begin{tabular}{lccc}
 \hline \hline
 & $\delta<\delta_{-} \quad$ & $\quad \delta_{-}<\delta<\delta_{+} \quad$ & $\quad \delta>\delta_{+}$ \\
 \hline
 $\text{sgn}[\varsigma^{(\text{I})}]$ & $-$ & $+$ & $+$  \\
 $\text{sgn}[\varsigma^{(\text{II})}]$ & $+$ & $+$ & $-$  \\
 \hline \hline
\end{tabular}
\end{table}

\begin{figure}[!]
    \centering
    \includegraphics[width=0.7\textwidth]{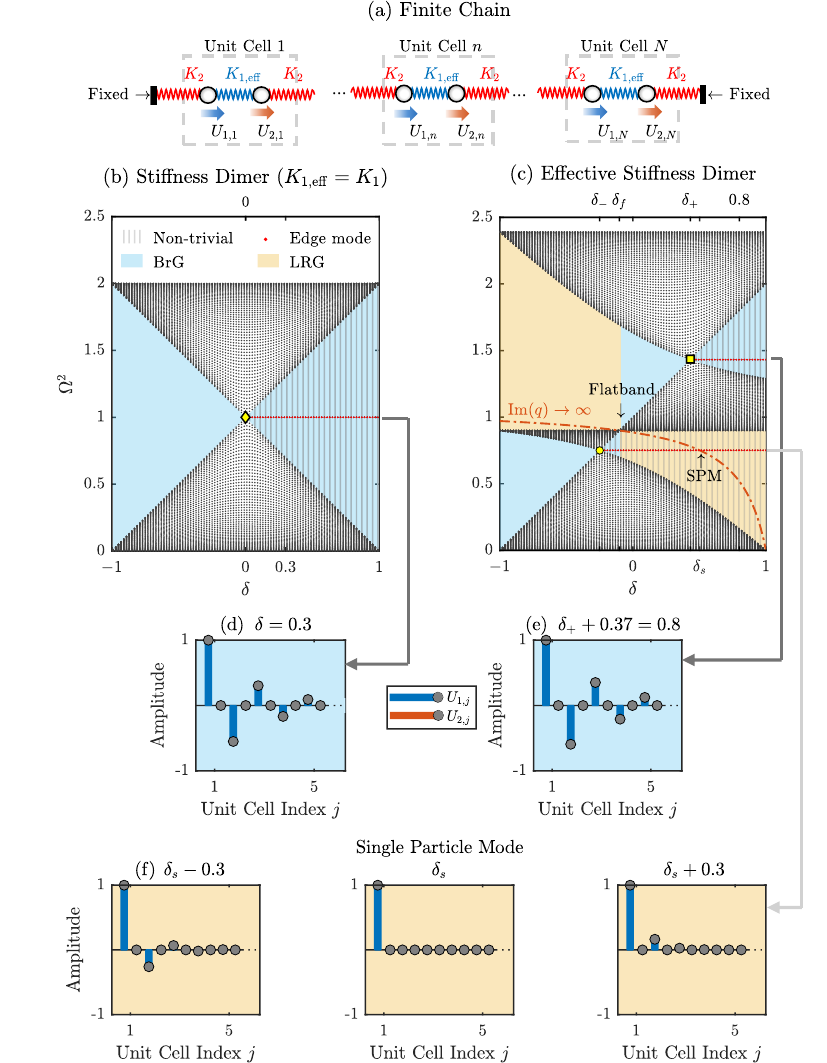}
    \caption{\textbf{Finite-chain eigenspectra and edge states\rev{, including the single-particle mode}.}
    (a) Schematic of finite chain with fixed boundaries.
    (b) Standard dimer spectrum; edge modes (red markers) emerge for $\delta > 0$.
    (c) Effective stiffness dimer spectrum; orange curve marks $K_{1,\text{eff}}=0$. Vertical lines: $\delta_-$, $\delta_f$, \pro{$\delta_+$, and $\delta_s$ (in ascending order for these parameters)}.
    (d) Edge mode shape in standard dimer at $\delta=0.3$.
    (e) Edge mode in upper BrG at $\delta=0.37$.
    (f) Edge modes in lower LRG at $\delta_s-0.3$, $\delta_s$, and $\delta_s+0.3$; center shows SPM (IPR$=1$).
    Parameters: $K=1$, $M=1$, $k_3=0.43$, $m_3=0.2$, $\lambda=1$, $N=60$.
}
\label{fig:consolidatedspectrum}
\end{figure}

\subsection{Finite chain spectrum and edge state localization\rev{, including the single-particle mode}}

We now analyze the eigenfrequency spectrum of a finite chain composed of $N=60$ unit cells with fixed boundary conditions [Fig.~\ref{fig:consolidatedspectrum}(a)]. This analysis allows for the direct observation of topological edge states that emerge within the bulk band gaps and provides a platform to verify the bulk-boundary correspondence.

As a baseline, Fig.~\ref{fig:consolidatedspectrum}(b) shows the spectrum for a standard stiffness dimer (without local resonators, \textit{i.e.}, $K_{1,\text{eff}} = K_1$) as a function of $\delta$. The bulk modes (black markers) form continuous bands separated by a single BrG (blue shaded region). This gap closes at $\delta=0$ (the Dirac point) and reopens into a topologically distinct phase, giving rise to a mid-gap edge state (red markers) for $\delta > 0$. The mode shape of this edge state, shown in Fig.~\ref{fig:consolidatedspectrum}(d), exhibits the characteristic exponentially decaying profile with alternating signs between adjacent  {unit-cells}, a hallmark of SSH-type topological modes.

For the intracell-resonant effective stiffness dimer chain, we solve the full finite eigenvalue problem. The equation of motion for the finite system is:
\begin{equation}
    (\mathbf{D}_{3N\times 3N}-\omega^2\mathbf{I}_{3N\times3N})\mathbf{u}_{3N\times1}=\mathbf{0}_{3N\times1},
    \label{eq:finite_eigenvalue}
\end{equation}
where the state vector is $\mathbf{u}_{3N\times1} = [U_{1,1}, \dots, U_{1,N}, U_{2,1}, \dots, U_{2,N}, v_{3,1}, \dots, v_{3,N}]^{T}$. The finite dynamical matrix is constructed as $\mathbf{D}_{3N\times 3N}=\mathbf{M}_{3N\times 3N}^{-1}\mathbf{K}_{3N\times 3N}$. The stiffness and mass matrices are given by:
\begin{equation}
    \mathbf{K}_{3N\times 3N} = \begin{pmatrix}
        \mathbf{K}_{11} & \mathbf{K}_{12} & \mathbf{K}_{13} \\
        \mathbf{K}_{12}^T & \mathbf{K}_{22} & \mathbf{K}_{23} \\
        \mathbf{K}_{13}^T & \mathbf{K}_{23}^T & \mathbf{K}_{33}
    \end{pmatrix}, \quad
    \mathbf{M}_{3N\times 3N} =\begin{pmatrix}
        \mathbf{M}_{11} & \mathbf{0} & \mathbf{0}\\
        \mathbf{0} & \mathbf{M}_{22} & \mathbf{0} \\
         \mathbf{0} & \mathbf{0} & \mathbf{M}_{33}
    \end{pmatrix},
\end{equation}
where the sub-matrices are defined as:
\begin{gather}
    \mathbf{K}_{11} = \mathbf{K}_{22} =\left(K_1+K_2+\frac{k_3 \lambda^2}{2}\right)\mathbf{I}_{N\times N}, \nonumber \\
    \mathbf{K}_{12} = \begin{pmatrix}
        \mathbf{0}_{1\times (N-1)} & 0\\
        -K_2  \mathbf{I}_{(N-1)\times (N-1)} & \mathbf{0}_{(N-1)\times 1}
    \end{pmatrix}
        -\left(K_1+\frac{k_3\lambda^2}{2}\right) \mathbf{I}_{N\times N}, \\
    \mathbf{K}_{13}=
    - \mathbf{K}_{23} = -k_3 \lambda\mathbf{I}_{N\times N}, \quad \mathbf{K}_{33} = 2k_3 \mathbf{I}_{N\times N}, \nonumber \\
    \mathbf{M}_{11} = \mathbf{M}_{22} = M \mathbf{I}_{N\times N}, \quad \mathbf{M}_{33} = 2m_3\mathbf{I}_{N\times N}. \nonumber
\end{gather}
Here, $\mathbf{I}_{N\times N}$ is the identity matrix and $\mathbf{0}$ represents zero matrices of appropriate dimensions.

The locally resonant (effective) stiffness dimer exhibits a richer spectrum, as shown in Fig.~\ref{fig:consolidatedspectrum}(c). We distinguish the band gap character by overlaying the frequency of the attenuation singularity (orange dashed curve), which marks the condition $K_{1,\text{eff}}=0$. The band gap containing this singularity is the LRG (yellow region), while the gap without it is the BrG (blue region). The BrG-to-LRG switching event is clearly visible at $\delta = \delta_f$.

Consistent with the bulk topological analysis, topological edge states (red markers) emerge in each band gap once $\delta$ crosses the corresponding Dirac point. The lower gap (I) hosts an edge state for $\delta > \delta_{-}$, and the upper gap (II) hosts one for $\delta > \delta_{+}$. Figure~\ref{fig:consolidatedspectrum}(e) displays the mode shape for an edge state in the upper BrG (gap II). It shows localization similar to the standard stiffness dimer [Fig.~\ref{fig:consolidatedspectrum}(d)], with exponential decay and multi-site participation, specifically characterized by out-of-phase motion between alternating sites.

\rev{The lower LRG enables a novel localization phenomenon.} When the frequency of the attenuation singularity, $\Omega_s$, \rev{coincides with} the frequency of the lower topological edge state, $\Omega_{\text{edge,I}}$, \rev{the SPM} condition is met. This occurs at the specific value $\delta = \delta_s$ {(see the Supplemental Material)}:
\begin{equation}
    \delta_{s} = \left(\frac{\kappa_3\lambda^2}{2} +\Omega_r^2\right)-\sqrt{\left(\frac{\kappa_3\lambda^2}{2} +\Omega_r^2-1\right)^2+ 2\kappa_3\lambda^2}.
    \label{eq:delta_s}
\end{equation}
At this precise point, the topological state achieves extreme localization, as illustrated in Fig.~\ref{fig:consolidatedspectrum}(f). The central panel of Fig.~\ref{fig:consolidatedspectrum}(f) ($\delta=\delta_s$) shows the mode shape confined entirely to a single primary mass at the chain's boundary, \textit{i.e.}, an SPM. \pro{This contrasts with the singular edge mode of Ref.~\cite{JANG2025}, which is confined to a single unit cell: here, the confinement is sub-unit-cell, collapsing onto one particle.}
\rev{Across this point, a} transition in the edge mode profile is observed: for $\delta < \delta_s$, the alternating sites oscillate out of phase \rev{as seen in the left panel of Fig.~\ref{fig:consolidatedspectrum}(f)}, whereas for $\delta > \delta_s$, they oscillate in phase \rev{as seen in the right panel of Fig.~\ref{fig:consolidatedspectrum}(f)}. Consequently, the SPM condition at $\delta=\delta_s$ serves as a singular crossover point between these distinct localization regimes.
We quantify this localization using the Inverse Participation Ratio (IPR):
\begin{equation}
\text{IPR}=\frac{\sum_{j=1}^{N}(U_{1,j}^4+U_{2,j}^4)}{\left(\sum_{j=1}^{N}(U_{1,j}^2+U_{2,j}^2)\right)^2}.
\label{eq:IPR}
\end{equation}
For the SPM at $\delta = \delta_s$, the IPR \pro{(computed over the primary masses)} is exactly unity\rev{---the theoretical maximum for localization in a discrete lattice---for a single boundary mode localized at either edge. This is reminiscent of the full dimerized limit of an SSH dimer~\cite{asboth2016short}, but here it occurs at a finite frequency within an LRG.}
\rev{This extreme confinement arises from three cooperating ingredients. The bulk topology (Zak phase) guarantees the edge mode and fixes its SSH-like nodal structure, placing zero amplitude on alternating particles; spectral isolation from the bulk bands sets its decay length; and the vanishing effective stiffness ($K_{1,\text{eff}} \to 0$) decouples the boundary unit cell from the rest of the chain---the local resonators precisely counterbalancing the inertial force of the boundary mass---collapsing the mode onto a single particle. Of these, only the existence of the edge mode is topologically guaranteed; the single-particle confinement follows from matching the edge-state frequency to $K_{1,\text{eff}}=0$ and is therefore driven by local resonance rather than protected by topology.}
\rev{It is important to emphasize here the necessity of the locally resonant mechanism for such extreme mode localization. In a Bragg gap, unless one of the springs is driven to zero stiffness, the imaginary wave number remains finite. In locally resonant systems (whether mass or stiffness dimers), as mentioned, the attenuation can become infinite in an idealized, non-dissipative system.}

\rev{This exact $\mathrm{IPR}=1$ value is a property of a small-amplitude linear eigenstate. Because the IPR characterizes the normalized modal profile rather than the absolute vibration amplitude, the state implies no large displacement, strain, or internal force, and the effective-stiffness description remains self-consistent in the infinitesimal-amplitude limit. It is therefore an idealized linear limit: in practice, geometric, material, or inertial nonlinearities and fabrication imperfections may shift the exact $K_{1,\mathrm{eff}}=0$ condition, but it nonetheless provides a clear design target for approaching extremely localized edge states.}

\rev{\pro{We also emphasize the value of} locating the edge state in the LRG \pro{from a design perspective: it frees the prescription of edge-mode frequencies from other structural constraints}. Locally resonant band gaps have been suggested to have particular value in creating forbidden regions at extremely low frequencies, below the first Bragg gap of the main chain. Contrasting against systems that exhibit pure Bragg gaps, we highlight two scenarios. First, consider a scenario where a specific mass budget is available for a system and a minimum quasi-static stiffness of the system is required. In a monoatomic or diatomic spring-mass chain, these parameters dictate the frequency of the Bragg gap. If one then adds local resonators, such as in a negative effective mass model~\cite{huang2009negative}, then by allocating mass to the local resonator an extremely low-frequency band gap can be enabled via a small stiffness of the local resonator. Second, imagine a scenario where a high long-wavelength sound speed is required. As highlighted in the Supplemental Material, in our locally resonant stiffness dimer system, the long-wavelength sound speed is dictated by the main chain mass, stiffness, and unit cell length. These same parameters pin the Bragg gap at high frequencies, whereas the parameters of the local resonator can be tuned to enable an arbitrarily low-frequency band gap.}

\begin{figure}[!]
    \centering
    \includegraphics[width=0.8\textwidth]{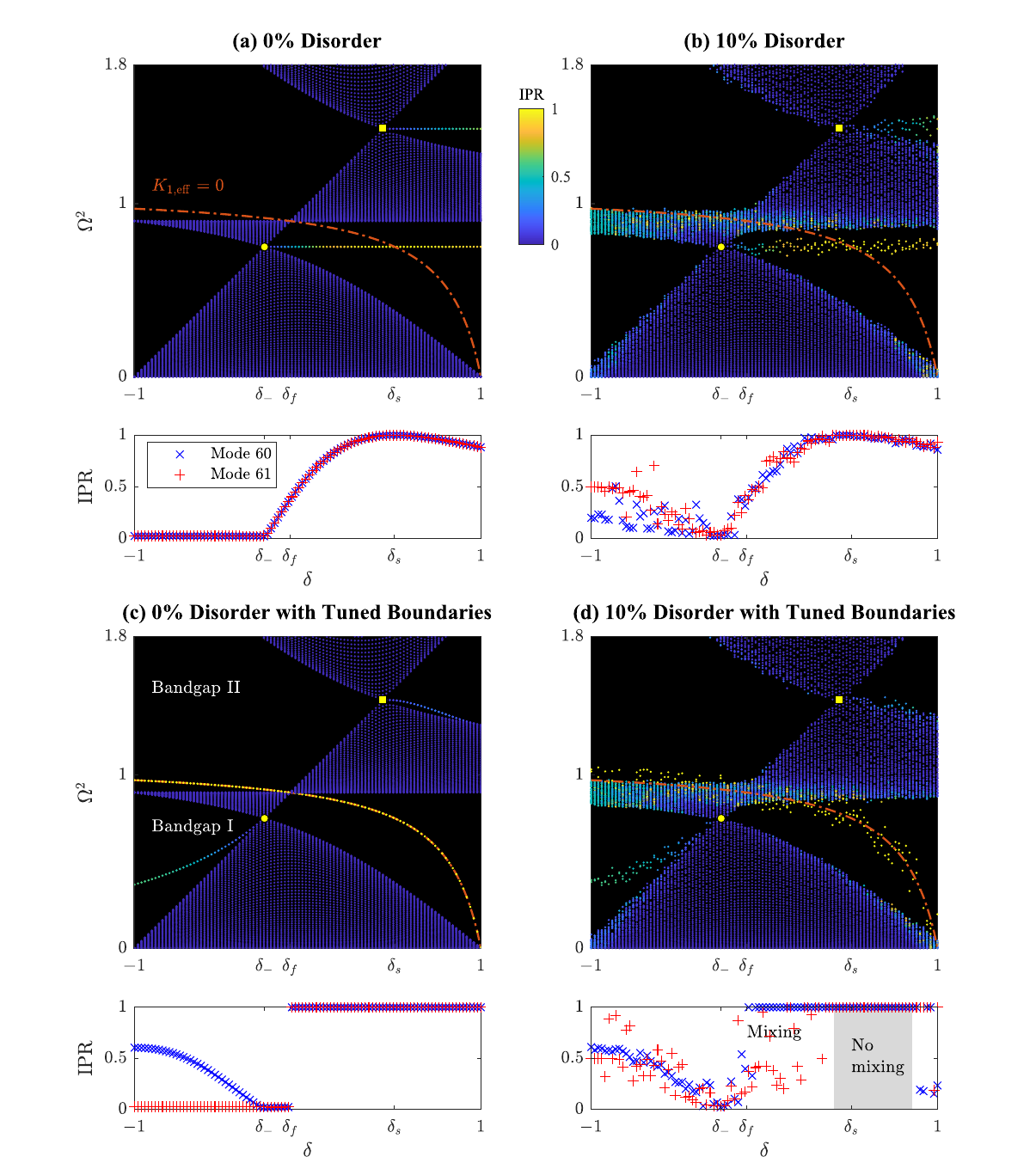}
 \caption{\textbf{Localization and disorder with tuned boundaries.}
    IPR shown by color (yellow = 1, blue = minimum). Lower panels: IPR for modes 60 and 61 (LRG edge modes).
    (a) No disorder: IPR $\to 1$ at $\delta_s$.
    (b) 10\% disorder: frequencies scatter but high IPR maintained near $\delta_s$.
    (c) Tuned boundaries, no disorder: edge modes track $K_{1,\text{eff}}=0$, achieving IPR $= 1$ broadly.
    (d) Tuned boundaries, 10\% disorder: \rev{SPM} in ``No mixing'' region; hybridization in ``Mixing'' region.
}
\label{fig:disorder}
\end{figure}

\subsection{Disorder analysis and tuned boundaries}
Having established the SPM at $\delta = \delta_s$, we now address two practical questions: (i)~how robust is this extreme localization to structural disorder, and (ii)~can the single-particle confinement be extended from an isolated fine-tuned point to a continuous parameter range?
These questions are central to any experimental realization\rev{: although the bulk topology guarantees the edge mode, the IPR$=1$ confinement requires matching its frequency to $K_{1,\text{eff}}=0$, so it is sensitive to detuning by the imperfections present in any real structure}. We address both through a disorder analysis and a boundary-tuning strategy, showing that \rev{boundary tuning} not only broadens the SPM into a finite parameter window but also \rev{improves its resilience to random perturbations by spectrally isolating the mode from the bulk bands}.

\rev{We first assess the effect of random disorder on the SPM, assigning} independent random variations to every stiffness element in the finite chain ($K_1$, $K_2$, and $k_3$) according to the following scheme:
\begin{align}
K_{1,j} &= K(1-\delta) + \Delta \cdot K_r^{(1,j)}, \label{eq:disorder_K1}\\
K_{2,j} &= K(1+\delta) + \Delta \cdot K_r^{(2,j)}, \label{eq:disorder_K2}\\
k_{3,j} &= k_3 \left(1 + \frac{\Delta \cdot K_r^{(3,j)}}{K}\right), \label{eq:disorder_k3}
\end{align}
where $K_r^{(\cdot,j)}$ are random numbers drawn from a uniform distribution over the interval $(-K, K)$, and $\Delta \in [0, 1)$ represents the disorder strength (\textit{e.g.}, $\Delta=0.1$ corresponds to 10\% disorder).

Figure~\ref{fig:disorder} presents a comparative study of localization, quantified by the IPR, across different configurations. Figure~\ref{fig:disorder}(a) serves as the reference case (no disorder), showing the sharp peak in IPR reaching exactly unity at $\delta=\delta_s$ {for mode numbers $60$ and $61$ (where the modes are arranged in an ascending order of eigenfrequencies}). When 10\% disorder is introduced [Fig.~\ref{fig:disorder}(b)], the translational symmetry is broken, and the eigenfrequencies of the edge modes scatter around the theoretical prediction. While the perfect IPR=1 condition is perturbed, the edge mode shows a degree of intrinsic robustness as it retains high localization values near $\delta_s$.

To overcome the limitation of the SPM being created at a single design point $\delta_s$, and to realize it across a broader range of system configurations, we propose a ``tuned boundary" strategy. The physical intuition is to enforce the \rev{SPM} condition locally at the boundaries, regardless of the bulk parameters.
We achieve this by modifying the stiffness of the boundary springs ($K_{2,1}$ at the left edge and $K_{2,N+1}$ at the right edge) such that the natural frequency of the isolated boundary mass matches the frequency where the bulk effective stiffness vanishes (\rev{\textit{i.e.}, $\omega_s$}).
This requires satisfying the condition $K_{2,\text{boundary}}/M = \omega_s^2$, \rev{where $K_{2,\text{boundary}}$ denotes the stiffness of either boundary spring. This condition} yields the closed-form tuning expressions:
\begin{equation}
K_{2,1}=\frac{Mk_{3,1}}{m_3\left( 1+\frac{\lambda^2k_{3,1}}{2K_{1,1}}\right)}, \quad
K_{2,N+1}=\frac{M k_{3,N}}{m_3\left( 1+\frac{\lambda^2 k_{3,N}}{2K_{1,N}}\right)}.
\label{eq:tuned_boundaries}
\end{equation}
This tuning effectively ``pins" the edge mode to the attenuation singularity curve ($K_{1,\text{eff}}=0$) across the parameter space. We note that this creates a ``defective'' boundary which does not support a topological edge mode in a strict sense; however, we suggest this is of interest regarding preservation of the single-particle localized edge mode. Previously, Yousef \textit{et al.}~\cite{yousef2024blueprint} have tuned eigenfrequencies of edge modes in a BrG by tuning the edges.
\rev{To the best of our knowledge, such} a study in LRGs or its impact on localization remains unexplored.
The efficacy of this strategy is demonstrated in Fig.~\ref{fig:disorder}(c). With tuned boundaries (and no disorder), the edge mode frequency perfectly tracks the orange dashed line corresponding to the attenuation singularity. Consequently, the edge mode maintains an IPR of exactly 1 over a broad range of $\delta$, extending the SPM from a point-like phenomenon to a stable phase across a parameter space.
Note that this tuning modifies the boundary conditions sufficiently to close the gap for BrG-based edge modes, which disappear, while new edge modes appear in gap I for $\delta < \delta_-$.

\rev{This boundary pinning also protects the SPM against disorder, provided the mode stays spectrally isolated from the bulk.}
Figure~\ref{fig:disorder}(d) shows the tuned system subject to 10\% disorder\rev{, revealing} two distinct regimes. In the ``Mixing" region, where the edge-mode frequency (tracking $K_{1,\text{eff}}=0$) approaches the bulk bands, disorder \rev{drives} hybridization between the edge state and bulk modes, leading to delocalization.
\rev{In the ``No mixing'' region, by contrast,} where the attenuation singularity is well separated from the bulk spectrum, the tuned SPM \rev{is markedly resilient: the IPR stays} close to unity and the mode remains spectrally distinct despite significant structural randomness. \rev{This resilience is a direct consequence of spectral isolation---keeping the edge mode far from the bulk bands suppresses disorder-induced hybridization.} \rev{The combination of spectral separation and boundary tuning thus provides a practical route to ultra-localized states in imperfect metamaterials whose edge modes originate from band topology.}

\section{Conclusion}

In this work, we introduced a locally resonant stiffness dimer as a framework for generating topological edge states within LRGs.
By mapping the lattice to an effective medium, we identified the attenuation singularity---where the effective stiffness vanishes---as the \pro{operational} signature of the LRG. \rev{We then showed that this singularity can be controllably migrated between adjacent band gaps by tuning a dimerization parameter, \pro{transferring the dominant local-resonance character---and hence the LRG designation---to the host gap}. \pro{Crucially, flat-band formation mediates this migration, with a `mixed' BrG--LRG gap appearing at the transition point.} Overall, this process preserves the topological invariant (the Zak phase): the system inherits non-trivial topology from a BrG without the gap closure usually required for topological transitions in LRGs.}

\rev{A central finding herein is the single-particle mode (SPM). When a topological edge state spectrally intersects the local-resonance-induced attenuation singularity, the vibrational energy collapses onto a single boundary particle, reaching an IPR$=1$---the theoretical limit for localization in a discrete lattice. \pro{This sharpens the single-unit-cell (`singular') edge mode of Ref.~\cite{JANG2025} to its ultimate sub-unit-cell limit.} We posit that the band topology guarantees the edge mode, while the vanishing effective stiffness drives its single-particle confinement, dynamically decoupling the boundary mass from the bulk as the local resonators counterbalance its inertia. This ideal IPR$=1$ is a small-amplitude linear limit and thus a concrete design target for extreme localization.}
\rev{In addition to enabling the SPM, we suggest that the identification of such edge modes in locally resonant metamaterials may have advantages when low-frequency operation is required in the face of other requirements, such as quasi-static stiffness.} \rev{By achieving the SPM below the first Bragg gap\pro{---and potentially even in the sub-Bragg regime---}we demonstrate an edge mode that inherits two advantages of LRGs: low frequency and unbounded attenuation.}

Finally, we addressed \rev{practical realizability} through boundary tuning. \rev{By matching the boundary resonance to the bulk singularity, the SPM can be ``pinned'' to the zero-effective-stiffness condition over a continuous parameter range rather than at a single fine-tuned point.}
\rev{These tuned states exhibit remarkable resilience against random disorder, provided they remain spectrally separated from the bulk bands.}

\pro{Beyond the theoretical framework presented here, the proposed system can be realized experimentally by combining two experimentally demonstrated components: four-bar-linkage cells and out-of-plane local resonators. For example, four-bar-linkage cells were fabricated and dynamically tested for longitudinal stress-wave propagation~\cite{zhou2024quasi,taniker2017generating}, and double-lever resonators have been experimentally verified to produce out-of-plane local resonance in locally resonant metamaterials~\cite{pires2025novel}. The main practical challenges are the integration of these components and the control of experimental imperfections, mainly damping and assembly errors~\cite{zhou2024quasi,taniker2017generating,pires2025novel}.}

\rev{More broadly, the underlying physics---a frequency-dependent effective parameter shaped by local resonance---is not specific to the truss-based unit cell and has been experimentally realized in acoustic Helmholtz-resonator metamaterials~\cite{fang2006ultrasonic}, single-phase elastic chiral metamaterials~\cite{zhu2014negative}, and electromagnetic split-ring (LC-)resonator metamaterials~\cite{smith2000composite}. The discrete lumped-mass realization is what makes the single-particle ($\mathrm{IPR}=1$) limit well-defined, but the more general singularity-migration phenomenon can be sought across these platforms.}

\rev{Looking ahead, extending this single-particle localization to prescribed locations within the bulk, and to two-dimensional lattices for reconfigurable waveguiding, may be promising directions for future work.}

\section*{ACKNOWLEDGMENTS}
G.S.S. acknowledges financial support through the Prime Minister’s Research Fellowship (PMRF ID: 0202572). R.C. gratefully acknowledges support from the Indian Institute of Science Startup Grant. K.Q. and N.B. acknowledge support from the U.S. Army Research Office (Grant No. W911NF-20-2-0182). I.F. acknowledges support from the Department of Defense (DoD) through the National Defense Science $\&$ Engineering Graduate (NDSEG) Fellowship Program. \pro{Finally, G.S.S. gratefully acknowledges the support and encouragement from his late mother, Mrs. Lakshmi, during the early stages of this work.}

\rev{\section*{Conflicts of interest}

The authors declare the following competing financial interest(s): N.B. is a founder and holds equity positions in Euler Materials.}

\section*{Author Contributions}
G.S.S., K.Q., and I.F. performed theoretical and numerical analyses.
G.T., N.B., and R.C. supervised the work.
All authors discussed the results and contributed to the manuscript.
G.S.S., K.Q., and I.F. contributed equally to this work.


\bibliography{Reference}

\clearpage
\newpage
\setcounter{figure}{0}
\renewcommand{\thefigure}{S\arabic{figure}}
\setcounter{table}{0}
\renewcommand{\thetable}{S\arabic{table}}
\setcounter{equation}{0}
\renewcommand{\theequation}{S\arabic{equation}}
\setcounter{page}{1}
\renewcommand{\thepage}{S\arabic{page}}
\setcounter{section}{0}
\setcounter{secnumdepth}{3}

\begin{center}
{\large\textbf{Supplemental Material:\\[0.4em]
Single-particle limit of a topological edge state in a locally resonant band gap}}
\end{center}
\vspace{1em}

\section{Equations of motion and dynamical matrices}
\label{sec:eom}

In this section, we derive the equations of motion and the corresponding dynamical matrices for the intracell-resonant stiffness dimer chain studied in the main text, as well as for the intercell-resonant configuration included here for completeness.

\subsection{Intracell-resonant stiffness dimer chain}
\label{sec:eom_intracell}

\begin{figure}[h!]
    \centering
    \includegraphics[width=8.6cm]{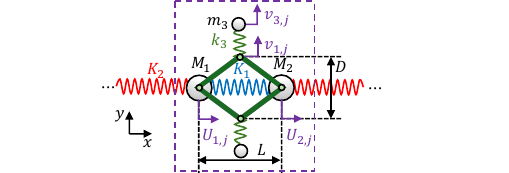}
    \caption{Unit cell configuration of the intracell-resonant stiffness dimer chain. Primary masses $M$ undergo horizontal displacements $U_1$ and $U_2$, while resonator masses $m_3$ undergo vertical displacement $v_3$. The truss geometry is characterized by parameters $L$ and $D$, with $\lambda = L/D$.}
    \label{fig:intracellunitcell}
\end{figure}

Consider an infinite quasi-one-dimensional stiffness dimer chain with local resonators attached to the intracell springs via a symmetric truss mechanism, as depicted in Fig.~\ref{fig:intracellunitcell}. Each unit cell contains two primary masses $M$ (with $M_1 = M_2 = M$) constrained to horizontal motion, alternating springs $K_1$ (intracell) and $K_2$ (intercell), and a pair of identical local resonators (mass $m_3$, spring $k_3$) coupled through hinged, massless truss members.

The equations of motion for the $j$th unit cell are:
\begin{align}
    \label{eq:eomU1_a}
    M \ddot{U}_{1,j} + K_1(U_{1,j}-U_{2,j}) + K_2(U_{1,j}-U_{2,j-1}) - 2k_3(v_{3,j}-v_{1,j})\left(\frac{L}{2D}\right) &= 0, \\
    \label{eq:eomU2_a}
    M \ddot{U}_{2,j} + K_1(U_{2,j}-U_{1,j}) + K_2(U_{2,j}-U_{1,j+1}) + 2k_3(v_{3,j}-v_{1,j})\left(\frac{L}{2D}\right) &= 0, \\
    \label{eq:eomv3_a}
    m_3 \ddot{v}_{3,j} + k_3(v_{3,j}-v_{1,j}) &= 0,
\end{align}
where $U_{1,j}$ and $U_{2,j}$ are the horizontal displacements of the two primary masses, $v_{3,j}$ is the vertical displacement of the resonator mass, and $v_{1,j}$ is the vertical displacement of the truss endpoint. The geometric parameters $L$ and $D$ characterize the massless rigid truss members, and we define $\lambda = L/D$.

The geometric constraint imposed by the rigid truss members relates the vertical displacement $v_{1,j}$ to the relative horizontal displacement $\Delta U_j = U_{2,j} - U_{1,j}$. Due to the symmetric two-bar geometry, this relative displacement is equally shared by the two inclined members, so each half-bar experiences $\Delta U_j/2$. Because the truss members maintain constant length, this relationship must satisfy:
\begin{equation}
    \left(\frac{L}{2}+\frac{\Delta U_{j}}{2}\right)^2 + \left(\frac{D}{2}+v_{1,j}\right)^2 = \left(\frac{L}{2}\right)^2 + \left(\frac{D}{2}\right)^2,
    \label{eq:v1_constraint}
\end{equation}
which yields:
\begin{equation}
    v_{1,j} = \frac{1}{2}\left(-D \pm \sqrt{D^2 - 2L\Delta U_{j} - (\Delta U_{j})^2}\right).
    \label{eq:v1_exact}
\end{equation}
We select the solution with the positive sign to ensure $v_{1,j} = 0$ when $\Delta U_{j} = 0$. For small displacements ($|\Delta U_{j}| \ll D, L$), we linearize by performing a Taylor expansion to first order:
\begin{equation}
    v_{1,j} \approx -\frac{\lambda}{2}(U_{2,j} - U_{1,j}).
    \label{eq:v1_linearized}
\end{equation}

Substituting Eq.~(\ref{eq:v1_linearized}) into Eqs.~(\ref{eq:eomU1_a})--(\ref{eq:eomv3_a}), we obtain the linearized equations of motion:
\begin{align}
    \label{eq:eomU1_b}
    M \ddot{U}_{1,j} + K_1(U_{1,j}-U_{2,j}) + K_2(U_{1,j}-U_{2,j-1}) - k_3\lambda\left[v_{3,j}+\frac{\lambda}{2}(U_{2,j}-U_{1,j})\right] &= 0, \\
    \label{eq:eomU2_b}
    M \ddot{U}_{2,j} + K_1(U_{2,j}-U_{1,j}) + K_2(U_{2,j}-U_{1,j+1}) + k_3\lambda\left[v_{3,j}+\frac{\lambda}{2}(U_{2,j}-U_{1,j})\right] &= 0, \\
    \label{eq:eomv3_b}
    m_3 \ddot{v}_{3,j} + k_3\left[v_{3,j}+\frac{\lambda}{2}(U_{2,j}-U_{1,j})\right] &= 0.
\end{align}
These are Eqs.~(1)--(3) in the main text.

Substituting Bloch wave solutions of the form $U_{1,j}(t) = U_1 e^{i(qj - \omega t)}$, $U_{2,j}(t) = U_2 e^{i(qj - \omega t)}$, and $v_{3,j}(t) = v_3 e^{i(qj - \omega t)}$ transforms the governing equations into the eigenvalue problem:
\begin{equation}
    \mathbf{D}(q)\mathbf{U} = \omega^2\mathbf{U},
    \label{eq:eigenvalue_problem}
\end{equation}
where the eigenvector is $\mathbf{U} = [U_1, U_2, \sqrt{2m_3/M}\,v_3]^T$ (the scaling of $v_3$ ensures a Hermitian dynamical matrix), and the $3\times3$ dynamical matrix is:
\begin{equation}
    \mathbf{D}(q) = 
    \begin{pmatrix} 
        \dfrac{K_1+K_2}{M}+\dfrac{k_3\lambda^2}{2M}
        & 
        -\left(\dfrac{K_1}{M}+\dfrac{K_2}{M}e^{-iq}+\dfrac{k_3\lambda^2}{2M}\right)
        & 
        -\dfrac{k_3 \lambda}{\sqrt{2Mm_3}} \\[2.5ex]
        -\left(\dfrac{K_1}{M}+\dfrac{K_2}{M}e^{iq}+\dfrac{k_3\lambda^2}{2M}\right)
        & 
        \dfrac{K_1+K_2}{M}+\dfrac{k_3\lambda^2}{2M}
        & 
        \dfrac{k_3 \lambda}{\sqrt{2Mm_3}} \\[2.5ex]
        -\dfrac{k_3\lambda}{\sqrt{2Mm_3}} & \dfrac{k_3\lambda}{\sqrt{2Mm_3}} & \dfrac{k_3}{m_3}
    \end{pmatrix}.
    \label{eq:D_intracell}
\end{equation}
Here, {$q=\tilde{\nu} d$ is the normalized wavenumber, where $d$ is the lattice constant and $\tilde{\nu}$ is the wavenumber}, and $\omega$ is the angular frequency. This is Eq.~(4) in the main text.

\subsection{Intercell-resonant stiffness dimer chain}
\label{sec:eom_intercell}

\begin{figure}[h!]
    \centering
    \includegraphics[width=8.6cm]{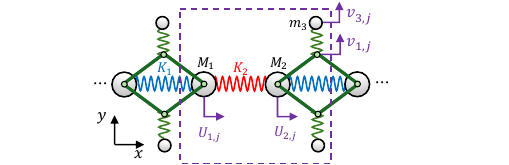}
    \caption{Unit cell configuration of the intercell-resonant stiffness dimer chain.}
    \label{fig:intercellunitcell}
\end{figure}

For completeness, we also consider the intercell-resonant configuration, where the local resonators are attached to the intercell springs $K_2$ rather than the intracell springs $K_1$, as depicted in Fig.~\ref{fig:intercellunitcell}.

The equations of motion for the $j$th unit cell are:
\begin{align}
    \label{eq:eomU1_intercell}
    M \ddot{U}_{1,j} + K_1(U_{1,j}-U_{2,j-1}) + K_2(U_{1,j}-U_{2,j}) + k_3\lambda(v_{3,j-1}-v_{1,j-1}) &= 0, \\
    \label{eq:eomU2_intercell}
    M \ddot{U}_{2,j} + K_1(U_{2,j}-U_{1,j+1}) + K_2(U_{2,j}-U_{1,j}) - k_3\lambda(v_{3,j}-v_{1,j}) &= 0, \\
    \label{eq:eomv3_intercell}
    m_3 \ddot{v}_{3,j} + k_3(v_{3,j}-v_{1,j}) &= 0.
\end{align}
Applying the geometric constraint and linearization yields:
\begin{equation}
    v_{1,j} \approx -\frac{\lambda}{2}(U_{1,j+1}-U_{2,j}).
    \label{eq:v1_intercell}
\end{equation}
Substituting this into the equations of motion gives the linearized form:
\begin{align}
    \label{eq:eomU1_intercell_lin}
    M \ddot{U}_{1,j} + K_1(U_{1,j}-U_{2,j-1}) + K_2(U_{1,j}-U_{2,j}) + k_3\lambda\left[v_{3,j-1}+\frac{\lambda}{2}(U_{1,j}-U_{2,j-1})\right] &= 0, \\
    \label{eq:eomU2_intercell_lin}
    M \ddot{U}_{2,j} + K_1(U_{2,j}-U_{1,j+1}) + K_2(U_{2,j}-U_{1,j}) - k_3\lambda\left[v_{3,j}+\frac{\lambda}{2}(U_{1,j+1}-U_{2,j})\right] &= 0, \\
    \label{eq:eomv3_intercell_lin}
    m_3 \ddot{v}_{3,j} + k_3\left[v_{3,j}+\frac{\lambda}{2}(U_{1,j+1}-U_{2,j})\right] &= 0.
\end{align}

Applying Bloch's theorem yields the eigenvalue problem $\mathbf{D}_{\text{intercell}}(q)\mathbf{U} = \omega^2\mathbf{U}$, where:
\begin{equation}
    \mathbf{D}_{\text{intercell}}(q) = 
    \begin{pmatrix} 
        \dfrac{K_1+K_2}{M}+\dfrac{k_3\lambda^2}{2M} 
        & 
        -\left(\dfrac{K_1}{M}e^{-iq}+\dfrac{K_2}{M}+\dfrac{k_3\lambda^2}{2M}e^{-iq}\right)
        & 
        \dfrac{k_3 e^{-iq}\lambda}{\sqrt{2Mm_3}} \\[2.5ex]
        -\left(\dfrac{K_1}{M}e^{iq}+\dfrac{K_2}{M}+\dfrac{k_3\lambda^2}{2M}e^{iq}\right)
        & 
        \dfrac{K_1+K_2}{M}+\dfrac{k_3\lambda^2}{2M} 
        & 
        -\dfrac{k_3 \lambda}{\sqrt{2Mm_3}} \\[2.5ex]
        \dfrac{k_3 e^{iq}\lambda}{\sqrt{2Mm_3}} & -\dfrac{k_3\lambda}{\sqrt{2Mm_3}} & \dfrac{k_3}{m_3}
    \end{pmatrix}.
    \label{eq:D_intercell}
\end{equation}

\section{Inversion symmetry analysis}
\label{sec:inversion}

In this section, we analyze the inversion symmetry of the dynamical matrices for both intracell- and intercell-resonant configurations. Inversion symmetry is a prerequisite for the quantization of the Zak phase in one-dimensional systems~\cite{jiao2021experimentally}, making this analysis essential for establishing the validity of topological invariants.

We seek a parity operator $\mathbf{P}$ that satisfies:
\begin{equation}
    \mathbf{P}\mathbf{D}(q)\mathbf{P}^{-1} = \mathbf{D}(-q).
    \label{eq:inversion_condition}
\end{equation}
Since $\mathbf{P}$ must be both Hermitian and unitary, it can be written in the general form:
\begin{equation}
    \mathbf{P} = 
    \begin{pmatrix}
        p_{11} & p_{12} & p_{13} \\
        \bar{p}_{12} & p_{22} & p_{23} \\
        \bar{p}_{13} & \bar{p}_{23} & p_{33}
    \end{pmatrix},
    \label{eq:P_general}
\end{equation}
where $p_{11}$, $p_{22}$, and $p_{33}$ are real, and the overbar denotes complex conjugation. Equation~(\ref{eq:inversion_condition}) can equivalently be written as:
\begin{equation}
    \mathbf{P}\mathbf{D}(q) = \mathbf{D}(-q)\mathbf{P}.
    \label{eq:inversion_commutation}
\end{equation}

\subsection{Intracell-resonant configuration}
\label{sec:inversion_intracell}

We now show that the intracell-resonant stiffness dimer chain possesses inversion symmetry, and therefore the Zak phase is a valid topological invariant.

Substituting $\mathbf{D}(q) = \mathbf{D}(q)$ from Eq.~(\ref{eq:D_intracell}) into Eq.~(\ref{eq:inversion_commutation}) and equating matrix elements on both sides yields a system of constraints on the elements of $\mathbf{P}$.

\textit{Step 1: Establishing that all elements are real.}
Equating the $(1,1)$ elements gives:
\begin{multline}
    p_{11}\left(\frac{K_1+K_2}{M}+\frac{k_3\lambda^2}{2M}\right)
    - p_{12}\left(\frac{K_1}{M}+\frac{K_2}{M}e^{iq}+\frac{k_3\lambda^2}{2M}\right)
    - p_{13}\frac{k_3\lambda}{\sqrt{2Mm_3}} \\
    = p_{11}\left(\frac{K_1+K_2}{M}+\frac{k_3\lambda^2}{2M}\right)
    - \bar{p}_{12}\left(\frac{K_1}{M}+\frac{K_2}{M}e^{iq}+\frac{k_3\lambda^2}{2M}\right)
    - \bar{p}_{13}\frac{k_3\lambda}{\sqrt{2Mm_3}}.
\end{multline}
This requires $p_{12} = \bar{p}_{12}$ and $p_{13} = \bar{p}_{13}$, implying both are real. Similarly, equating the $(2,2)$ elements shows $p_{23} = \bar{p}_{23}$, so $p_{23}$ is also real. Thus, all elements of $\mathbf{P}$ are real.

\textit{Step 2: Determining diagonal elements.}
Equating the $(1,2)$ elements and rearranging:
\begin{equation}
    (p_{11}-p_{22})\left(\frac{K_1}{M}+\frac{k_3\lambda^2}{2M}\right) + p_{11}\frac{K_2}{M}e^{-iq} - p_{22}\frac{K_2}{M}e^{iq}
    = (p_{13}+p_{23})\frac{k_3\lambda}{\sqrt{2Mm_3}}.
    \label{eq:constraint_12}
\end{equation}
We assume the parity operator is a geometric symmetry operator independent of specific parameter values. For Eq.~(\ref{eq:constraint_12}) to hold for all $q$, the terms proportional to $e^{\pm iq}$ must vanish independently, yielding:
\begin{equation}
    p_{11} = p_{22} = 0.
    \label{eq:p11_p22}
\end{equation}
Substituting back into Eq.~(\ref{eq:constraint_12}) gives:
\begin{equation}
    p_{13} + p_{23} = 0.
    \label{eq:p13_p23}
\end{equation}

\textit{Step 3: Determining off-diagonal elements.}
Equating the $(1,3)$ elements with $p_{11} = p_{22} = 0$ and $p_{23} = -p_{13}$:
\begin{equation}
    (p_{33}+p_{12})\frac{k_3\lambda}{\sqrt{2Mm_3}}
    = p_{13}\left(\frac{2K_1+K_2(1+e^{iq})+k_3\lambda^2}{M}-\frac{k_3}{m_3}\right).
    \label{eq:constraint_13}
\end{equation}
For this to hold independently of parameters, both sides must vanish:
\begin{equation}
    p_{33} + p_{12} = 0, \quad p_{13} = 0.
    \label{eq:p33_p12_p13}
\end{equation}

Combining Eqs.~(\ref{eq:p11_p22}), (\ref{eq:p13_p23}), and (\ref{eq:p33_p12_p13}), the parity operator takes the form:
\begin{equation}
    \mathbf{P} = 
    \begin{pmatrix}
        0 & -p_{33} & 0 \\
        -p_{33} & 0 & 0 \\
        0 & 0 & p_{33}
    \end{pmatrix}.
    \label{eq:P_intracell_form}
\end{equation}
The determinant is $|\mathbf{P}| = -p_{33}^3$. Since $\mathbf{P}$ is real and unitary, $|\mathbf{P}| = \pm 1$, which implies $p_{33} = \pm 1$.

For $p_{33} = 1$:
\begin{equation}
    \mathbf{P} = \mathbf{P}^{\dagger} = \mathbf{P}^{-1} = 
    \begin{pmatrix}
        0 & -1 & 0 \\
        -1 & 0 & 0 \\
        0 & 0 & 1
    \end{pmatrix}.
    \label{eq:P_intracell_final}
\end{equation}
This is the parity operator given in the main text [Eq.~(16)]. Physically, it exchanges the two primary masses ($U_1 \leftrightarrow -U_2$) while leaving the resonator coordinate unchanged, reflecting the geometric inversion center of the unit cell.

\textbf{Conclusion:} The dynamical matrix $\mathbf{D}(q)$ of the intracell-resonant stiffness dimer chain possesses inversion symmetry, and the Zak phase is a valid topological invariant.

\subsection{Intercell-resonant configuration}
\label{sec:inversion_intercell}

We now show that the intercell-resonant stiffness dimer chain does \textit{not} possess inversion symmetry in its full $3\times3$ representation, and therefore the Zak phase computed from the $3\times3$ eigenvectors is not quantized.

Substituting $\mathbf{D}(q) = \mathbf{D}_{\text{intercell}}(q)$ from Eq.~(\ref{eq:D_intercell}) into Eq.~(\ref{eq:inversion_commutation}) and following the same procedure as above, we first establish that all elements of $\mathbf{P}$ are real.

Equating the $(1,2)$ elements and rearranging:
\begin{equation}
    (p_{11}-p_{22})\left(\frac{K_1}{M}e^{iq}+\frac{K_2}{M}+\frac{k_3\lambda^2}{2M}e^{iq}\right)
    - p_{13}\frac{k_3\lambda}{\sqrt{2Mm_3}}
    = p_{23}\frac{k_3 e^{iq}\lambda}{\sqrt{2Mm_3}}.
    \label{eq:constraint_12_intercell}
\end{equation}
For this to hold for all $q$ and all parameter values:
\begin{equation}
    p_{11} = p_{22}, \quad p_{13} = p_{23} = 0.
    \label{eq:intercell_constraints_1}
\end{equation}

Equating the $(1,3)$ elements with these constraints:
\begin{equation}
    p_{11} e^{-iq} - p_{12} = p_{33} e^{iq}.
    \label{eq:constraint_13_intercell}
\end{equation}
For this to hold for all $q$:
\begin{equation}
    p_{11} = p_{12} = p_{33} = 0.
    \label{eq:intercell_constraints_2}
\end{equation}

Combining all constraints, the parity operator becomes:
\begin{equation}
    \mathbf{P} = 
    \begin{pmatrix}
        0 & 0 & 0 \\
        0 & 0 & 0 \\
        0 & 0 & 0
    \end{pmatrix}.
    \label{eq:P_intercell_null}
\end{equation}
Since a valid parity operator must be unitary (and hence non-singular), this null matrix cannot serve as a parity operator.

\textbf{Conclusion:} The dynamical matrix $\mathbf{D}_{\text{intercell}}(q)$ does not possess inversion symmetry in its $3\times3$ representation, and therefore the Zak phase computed from the full eigenvectors is not a valid topological invariant.

\subsection{Effective $2\times2$ framework}
\label{sec:inversion_effective}

Here we construct a reduced $2\times2$ effective model by eliminating the local resonator degree of freedom and analyze its inversion symmetry.

In the effective framework, $\mathbf{D}_{\text{Intracell}3\times3}(q)$ and $\mathbf{D}_{\text{Intercell}3\times3}(q)$ become
\begin{equation}
    \mathbf{D}_{\text{intracell,eff}}(q,\omega) = 
    \frac{1}{M}
    \begin{pmatrix} 
    K_{1,\text{eff}}(\omega)+K_2 & -\left(K_{1,\text{eff}}(\omega)+K_2 e^{-iq}\right) \\[1.5ex]
    -\left(K_{1,\text{eff}}(\omega)+K_2 e^{iq}\right) & K_{1,\text{eff}}(\omega)+K_2
    \end{pmatrix}
    \label{eq:D_intracell_eff}
\end{equation}
and
\begin{equation}
    \mathbf{D}_{\text{intercell,eff}}(q,\omega) = 
    \frac{1}{M}
    \begin{pmatrix} 
    K_{1,\text{eff}}(\omega)+K_2 & -\left(K_{1,\text{eff}}(\omega) e^{-iq}+K_2\right) \\[1.5ex]
    -\left(K_{1,\text{eff}}(\omega) e^{iq}+K_2\right) & K_{1,\text{eff}}(\omega)+K_2
    \end{pmatrix},
    \label{eq:D_intercell_eff}
\end{equation}
respectively, where
\begin{equation}
    K_{1,\text{eff}}(\omega) = K_1 + \frac{k_3\lambda^2}{2}\frac{\omega^2}{\omega^2-\omega_0^2}
\end{equation}
and $\omega_0 = \sqrt{k_3/m_3}$.

We explicitly verify that the Pauli matrix $\boldsymbol{\sigma}_x$,
\begin{equation}
    \boldsymbol{\sigma}_x = 
    \begin{pmatrix}
    0 & 1 \\
    1 & 0
    \end{pmatrix},
\end{equation}
acts as the parity operator $\mathbf{P}$ for both reduced models, satisfying Eq.~\eqref{eq:inversion_commutation}, i.e.,
\begin{equation}
    \boldsymbol{\sigma}_x \mathbf{D}_{\text{intracell,eff}}(q) = \mathbf{D}_{\text{intracell,eff}}(-q) \boldsymbol{\sigma}_x
\end{equation}
and
\begin{equation}
    \boldsymbol{\sigma}_x \mathbf{D}_{\text{intercell,eff}}(q) = \mathbf{D}_{\text{intercell,eff}}(-q) \boldsymbol{\sigma}_x.
\end{equation}
Therefore, the reduced $2\times2$ effective intercell model is inversion symmetric, although the full $3\times3$ system is not. This is one of the advantages of analyzing the system in an effective framework and only keeping track of the motion of the main masses.

\section{Analytical criteria for band-touching and flat band phenomena}
\label{sec:analytical}

In this section, we derive analytical expressions for the band-edge frequencies at high-symmetry points in the Brillouin zone and establish the conditions for band touching (Dirac points) and flat-band formation. Throughout this section, we consider the intracell-resonant stiffness dimer chain with $M_1 = M_2 = M$, $K_1 = K(1-\delta)$, and $K_2 = K(1+\delta)$, where $M > 0$, $K > 0$, $\lambda > 0$, $m_3 > 0$, $k_3 > 0$, and $\delta \in (-1,1)$.

With this parametrization, the dynamical matrix for the intercell configuration becomes:
\begin{equation}
    \mathbf{D}(q) = 
    \begin{pmatrix} 
        \dfrac{2K}{M}+\dfrac{k_3\lambda^2}{2M}
        & 
        -\left(\dfrac{K}{M}[1-\delta+(1+\delta)e^{-iq}]+\dfrac{k_3\lambda^2}{2M}\right)
        & 
        -\dfrac{k_3 \lambda}{\sqrt{2Mm_3}} \\[2.5ex]
        -\left(\dfrac{K}{M}[1-\delta+(1+\delta)e^{iq}]+\dfrac{k_3\lambda^2}{2M}\right)
        & 
        \dfrac{2K}{M}+\dfrac{k_3\lambda^2}{2M}
        & 
        \dfrac{k_3 \lambda}{\sqrt{2Mm_3}} \\[2.5ex]
        -\dfrac{k_3\lambda}{\sqrt{2Mm_3}} & \dfrac{k_3\lambda}{\sqrt{2Mm_3}} & \dfrac{k_3}{m_3}
    \end{pmatrix}.
    \label{eq:D_parametrized}
\end{equation}

\subsection{Band-edge frequencies at $q = 0$ (Brillouin zone center)}
\label{sec:bandedge_q0}

At the Brillouin zone center ($q = 0$), we have $e^{\pm iq} = 1$, and the dynamical matrix simplifies to:
\begin{equation}
    \mathbf{D}(0) = 
    \begin{pmatrix} 
        \dfrac{2K}{M}+\dfrac{k_3\lambda^2}{2M} & -\left(\dfrac{2K}{M}+\dfrac{k_3\lambda^2}{2M}\right) & -\dfrac{k_3 \lambda}{\sqrt{2Mm_3}} \\[2.5ex]
        -\left(\dfrac{2K}{M}+\dfrac{k_3\lambda^2}{2M}\right) & \dfrac{2K}{M}+\dfrac{k_3\lambda^2}{2M} & \dfrac{k_3 \lambda}{\sqrt{2Mm_3}} \\[2.5ex]
        -\dfrac{k_3\lambda}{\sqrt{2Mm_3}} & \dfrac{k_3\lambda}{\sqrt{2Mm_3}} & \dfrac{k_3}{m_3}
    \end{pmatrix}.
    \label{eq:D_q0}
\end{equation}
Note that this matrix is independent of $\delta$.

The eigenvalues $\omega_{0,j}^2$ (with $j \in \{1,2,3\}$) give the band-edge frequencies at the Brillouin zone center. Solving the characteristic equation $\det[\mathbf{D}(0) - \omega^2 \mathbf{I}] = 0$ yields:
\begin{equation}
    \omega_{0,1}^2 = 0,
    \label{eq:omega01}
\end{equation}
\begin{equation}
    \omega_{0,2}^2 = \frac{1}{2} \left[ \left(\frac{4K+k_3\lambda^2}{M} + \frac{k_3}{m_3}\right) - \sqrt{\left(\frac{4K+k_3\lambda^2}{M} + \frac{k_3}{m_3}\right)^2 - \frac{16Kk_3}{Mm_3}} \right],
    \label{eq:omega02}
\end{equation}
\begin{equation}
    \omega_{0,3}^2 = \frac{1}{2} \left[ \left(\frac{4K+k_3\lambda^2}{M} + \frac{k_3}{m_3}\right) + \sqrt{\left(\frac{4K+k_3\lambda^2}{M} + \frac{k_3}{m_3}\right)^2 - \frac{16Kk_3}{Mm_3}} \right].
    \label{eq:omega03}
\end{equation}

To verify that these eigenvalues are real and properly ordered, we examine the discriminant:
\begin{equation}
    \left(\frac{4K+k_3\lambda^2}{M} + \frac{k_3}{m_3}\right)^2 - \frac{16Kk_3}{Mm_3}
    = \left(\frac{4K+k_3\lambda^2}{M} - \frac{k_3}{m_3}\right)^2 + \frac{4k_3^2\lambda^2}{Mm_3} > 0.
    \label{eq:discriminant_q0}
\end{equation}
Since this is strictly positive, the square root is real. Furthermore:
\begin{equation}
    0 < \sqrt{\left(\frac{4K+k_3\lambda^2}{M} + \frac{k_3}{m_3}\right)^2 - \frac{16Kk_3}{Mm_3}} < \frac{4K+k_3\lambda^2}{M} + \frac{k_3}{m_3},
\end{equation}
which implies $\omega_{0,2}^2 > 0$ and $\omega_{0,3}^2 > \omega_{0,2}^2$. Therefore:
\begin{equation}
    \omega_{0,1}^2 < \omega_{0,2}^2 < \omega_{0,3}^2.
    \label{eq:ordering_q0}
\end{equation}

Introducing the non-dimensional parameters $\kappa_3 = k_3/(2K)$, $\Omega_r^2 = k_3 M/(2K m_3)$, and the non-dimensional frequency $\Omega = \omega\sqrt{M/(2K)}$, the band-edge frequencies become:
\begin{equation}
    \Omega_{0,1}^2 = 0,
    \label{eq:Omega01}
\end{equation}
\begin{equation}
    \Omega_{0,2}^2 = \left(1+\kappa_3\lambda^2 + \frac{\Omega_r^2}{2}\right) - \sqrt{\left(1+\kappa_3\lambda^2 + \frac{\Omega_r^2}{2}\right)^2 - 2\Omega_r^2},
    \label{eq:Omega02}
\end{equation}
\begin{equation}
    \Omega_{0,3}^2 = \left(1+\kappa_3\lambda^2 + \frac{\Omega_r^2}{2}\right) + \sqrt{\left(1+\kappa_3\lambda^2 + \frac{\Omega_r^2}{2}\right)^2 - 2\Omega_r^2}.
    \label{eq:Omega03}
\end{equation}
Importantly, these band-edge frequencies at $q = 0$ are independent of the dimerization parameter $\delta$.

\subsection{Band-edge frequencies at $q = \pi$ (Brillouin zone edge)}
\label{sec:bandedge_qpi}

At the Brillouin zone edge ($q = \pi$), we have $e^{\pm iq} = -1$, and the dynamical matrix becomes:
\begin{equation}
    \mathbf{D}(\pi) = 
    \begin{pmatrix} 
        \dfrac{2K}{M}+\dfrac{k_3\lambda^2}{2M} & \dfrac{2K\delta}{M}-\dfrac{k_3\lambda^2}{2M} & -\dfrac{k_3 \lambda}{\sqrt{2Mm_3}} \\[2.5ex]
        \dfrac{2K\delta}{M}-\dfrac{k_3\lambda^2}{2M} & \dfrac{2K}{M}+\dfrac{k_3\lambda^2}{2M} & \dfrac{k_3 \lambda}{\sqrt{2Mm_3}} \\[2.5ex]
        -\dfrac{k_3\lambda}{\sqrt{2Mm_3}} & \dfrac{k_3\lambda}{\sqrt{2Mm_3}} & \dfrac{k_3}{m_3}
    \end{pmatrix}.
    \label{eq:D_qpi}
\end{equation}
Note that this matrix depends on $\delta$.

The eigenvalues $\omega_{\pi,j}^2$ (with $j \in \{1,2,3\}$) give the band-edge frequencies at the Brillouin zone edge:
\begin{equation}
    \omega_{\pi,1}^2 = \frac{2K(1+\delta)}{M},
    \label{eq:omegapi1}
\end{equation}
\begin{equation}
    \omega_{\pi,2}^2 = \frac{1}{2} \left[ \left(\frac{2K(1-\delta)+k_3\lambda^2}{M} + \frac{k_3}{m_3}\right) -
    \sqrt{\left(\frac{2K(1-\delta)+k_3\lambda^2}{M} - \frac{k_3}{m_3}\right)^2 + \frac{4k_3^2\lambda^2}{Mm_3}} \right],
    \label{eq:omegapi2}
\end{equation}
\begin{equation}
    \omega_{\pi,3}^2 = \frac{1}{2} \left[ \left(\frac{2K(1-\delta)+k_3\lambda^2}{M} + \frac{k_3}{m_3}\right) +
    \sqrt{\left(\frac{2K(1-\delta)+k_3\lambda^2}{M} - \frac{k_3}{m_3}\right)^2 + \frac{4k_3^2\lambda^2}{Mm_3}} \right].
    \label{eq:omegapi3}
\end{equation}
From the structure of these expressions, we can conclude that $\omega_{\pi,2}^2 < \omega_{\pi,3}^2$, but the ordering of $\omega_{\pi,1}^2$ relative to the others depends on $\delta$.

In non-dimensional form:
\begin{equation}
    \Omega_{\pi,1}^2 = 1+\delta,
    \label{eq:Omegapi1}
\end{equation}
\begin{equation}
    \Omega_{\pi,2}^2 = \frac{1-\delta}{2}+\kappa_3\lambda^2+\frac{\Omega_r^2}{2} - \sqrt{\left(\frac{1-\delta}{2}+\kappa_3\lambda^2-\frac{\Omega_r^2}{2}\right)^2 + 2\kappa_3\lambda^2\Omega_r^2},
    \label{eq:Omegapi2}
\end{equation}
\begin{equation}
    \Omega_{\pi,3}^2 = \frac{1-\delta}{2}+\kappa_3\lambda^2+\frac{\Omega_r^2}{2} + \sqrt{\left(\frac{1-\delta}{2}+\kappa_3\lambda^2-\frac{\Omega_r^2}{2}\right)^2 + 2\kappa_3\lambda^2\Omega_r^2}.
    \label{eq:Omegapi3}
\end{equation}
All three band-edge frequencies at $q = \pi$ depend on $\delta$.

\subsection{Band touching at $q = \pi$ (Dirac points)}
\label{sec:band_touching}

We now derive the conditions for band touching at the Brillouin zone edge. Since the band-edge frequencies at $q = 0$ are independent of $\delta$, band touching can only occur at $q = \pi$.

First, we examine whether $\omega_{\pi,2}^2$ and $\omega_{\pi,3}^2$ can become degenerate. The discriminant under the square root in Eqs.~(\ref{eq:omegapi2}) and (\ref{eq:omegapi3}) is:
\begin{equation}
    \Delta_{\pi}(\delta) = \left(\frac{2K(1-\delta)+k_3\lambda^2}{M} - \frac{k_3}{m_3}\right)^2 + \frac{4k_3^2\lambda^2}{Mm_3}.
    \label{eq:discriminant_pi}
\end{equation}
Since $\Delta_{\pi}$ is a sum of a squared term and a strictly positive term, it is always positive and can never vanish. Therefore, $\omega_{\pi,2}^2$ and $\omega_{\pi,3}^2$ are never degenerate.

Consequently, band touching can occur only when the first eigenvalue $\omega_{\pi,1}^2$ equals one of the other two. This condition is satisfied when:
\begin{equation}
    \frac{2K(1+\delta)}{M} = \frac{1}{2} \left[ \left(\frac{2K(1-\delta)+k_3\lambda^2}{M} + \frac{k_3}{m_3}\right) \pm \sqrt{\Delta_{\pi}(\delta)} \right].
    \label{eq:touching_condition}
\end{equation}

Solving this equation yields the critical values of $\delta$ at which band touching occurs:
\begin{equation}
    \delta_{\pm} = \frac{1}{8K}\left(k_3\lambda^2+\frac{2k_3 M}{m_3} - 4K \pm \sqrt{\left(k_3\lambda^2+\frac{2k_3 M}{m_3} - 4K\right)^2 + 16 k_3\lambda^2 K}\right).
    \label{eq:delta_pm_dim}
\end{equation}

In non-dimensional form:
\begin{equation}
    \delta_{\pm} = \left(\frac{\kappa_3\lambda^2}{4}+\frac{\Omega_r^2-1}{2}\right) \pm \sqrt{\left(\frac{\kappa_3\lambda^2}{4}+\frac{\Omega_r^2-1}{2}\right)^2 + \frac{\kappa_3\lambda^2}{2}}.
    \label{eq:delta_pm_SM}
\end{equation}
This is Eq.~(19) in the main text.

From Eq.~(\ref{eq:delta_pm_SM}), we observe that:
\begin{itemize}
    \item The term under the square root is always positive, so $\delta_+$ and $\delta_-$ are always real.
    \item Since the square root is larger than the magnitude of the first term when that term is negative, we have $\delta_- < 0 < \delta_+$ for all physically relevant parameters.
\end{itemize}

The corresponding frequencies at which band touching occurs are:
\begin{equation}
    \omega_{\pm}^2 = \frac{2K(1+\delta_{\pm})}{M},
    \label{eq:omega_pm_dim}
\end{equation}
or in non-dimensional form:
\begin{equation}
    \Omega_{\pm}^2 = 1 + \delta_{\pm} = \frac{1}{2} + \frac{\kappa_3\lambda^2}{4} + \frac{\Omega_r^2}{2} \pm \sqrt{\left(\frac{\kappa_3\lambda^2}{4}+\frac{\Omega_r^2-1}{2}\right)^2 + \frac{\kappa_3\lambda^2}{2}}.
    \label{eq:Omega_pm}
\end{equation}

{Rearranging Eq.~\eqref{eq:Omega_pm}: 
\begin{equation}
    \Omega_{\pm}^2 = \frac{1}{2}\left[ \left(1+ \frac{\kappa_3\lambda^2}{2}+\Omega_r^2 \right)  \pm \sqrt{\left(1+\frac{\kappa_3\lambda^2}{2}+\Omega_r^2\right)^2 -4 \Omega_r^2} \right].
    \label{eq:Omega_pm2}
\end{equation}}

These are the Dirac point frequencies mentioned in the main text. DP1 occurs at $(\delta_-, \Omega_-^2)$ where the bottom and middle bands touch, and DP2 occurs at $(\delta_+, \Omega_+^2)$ where the middle and top bands touch.

\subsection{Flat-band formation condition}
\label{sec:flat_band}

A flat band occurs when one of the passbands becomes dispersionless, i.e., when the band-edge frequency at $q = 0$ coincides with a band-edge frequency at $q = \pi$. For the middle band (band 2) to become flat, we require:
\begin{equation}
    \omega_{0,2}^2 = \omega_{\pi,j}^2 \quad \text{for some } j \in \{1,2,3\}.
    \label{eq:flat_band_condition_general}
\end{equation}

We examine each case:

\textit{Case 1: $\omega_{0,2}^2 = \omega_{\pi,2}^2$.}
Equating Eqs.~(\ref{eq:omega02}) and (\ref{eq:omegapi2}) and solving for $\delta$ yields only $\delta = -1$, which is outside the valid range.

\textit{Case 2: $\omega_{0,2}^2 = \omega_{\pi,3}^2$.}
Similarly, this yields only $\delta = -1$.

\textit{Case 3: $\omega_{0,2}^2 = \omega_{\pi,1}^2$.}
This condition gives:
\begin{equation}
    \frac{1}{2} \left[ \left(\frac{4K+k_3\lambda^2}{M} + \frac{k_3}{m_3}\right) - \sqrt{\left(\frac{4K+k_3\lambda^2}{M} + \frac{k_3}{m_3}\right)^2 - \frac{16Kk_3}{Mm_3}} \right] = \frac{2K(1+\delta_f)}{M}.
    \label{eq:flat_band_condition}
\end{equation}

Solving for $\delta_f$:
\begin{equation}
    \delta_f = \frac{1}{4K} \left[ \left(k_3\lambda^2 + \frac{k_3 M}{m_3}\right) - \sqrt{\left(k_3\lambda^2 + \frac{k_3 M}{m_3} - 4K\right)^2 + 16 k_3\lambda^2 K} \right].
    \label{eq:delta_f_dim}
\end{equation}

In non-dimensional form:
\begin{equation}
    \delta_f = \frac{\kappa_3\lambda^2 + \Omega_r^2}{2} - \sqrt{\left(\frac{\kappa_3\lambda^2 + \Omega_r^2}{2} - 1\right)^2 + 2\kappa_3\lambda^2}.
    \label{eq:delta_f_SM}
\end{equation}
This is Eq.~(14) in the main text.

The flat-band frequency is $\Omega_{0,2}$ [Eq.~(\ref{eq:Omega02})], which equals $\Omega_{\pi,1} = \sqrt{1+\delta_f}$ at this condition.

\kq{\section{Bragg and local-resonance diagnostics for band gap classification}}
\label{sec:gap_type}

\begin{figure}
    \centering
    \includegraphics[width=\textwidth]{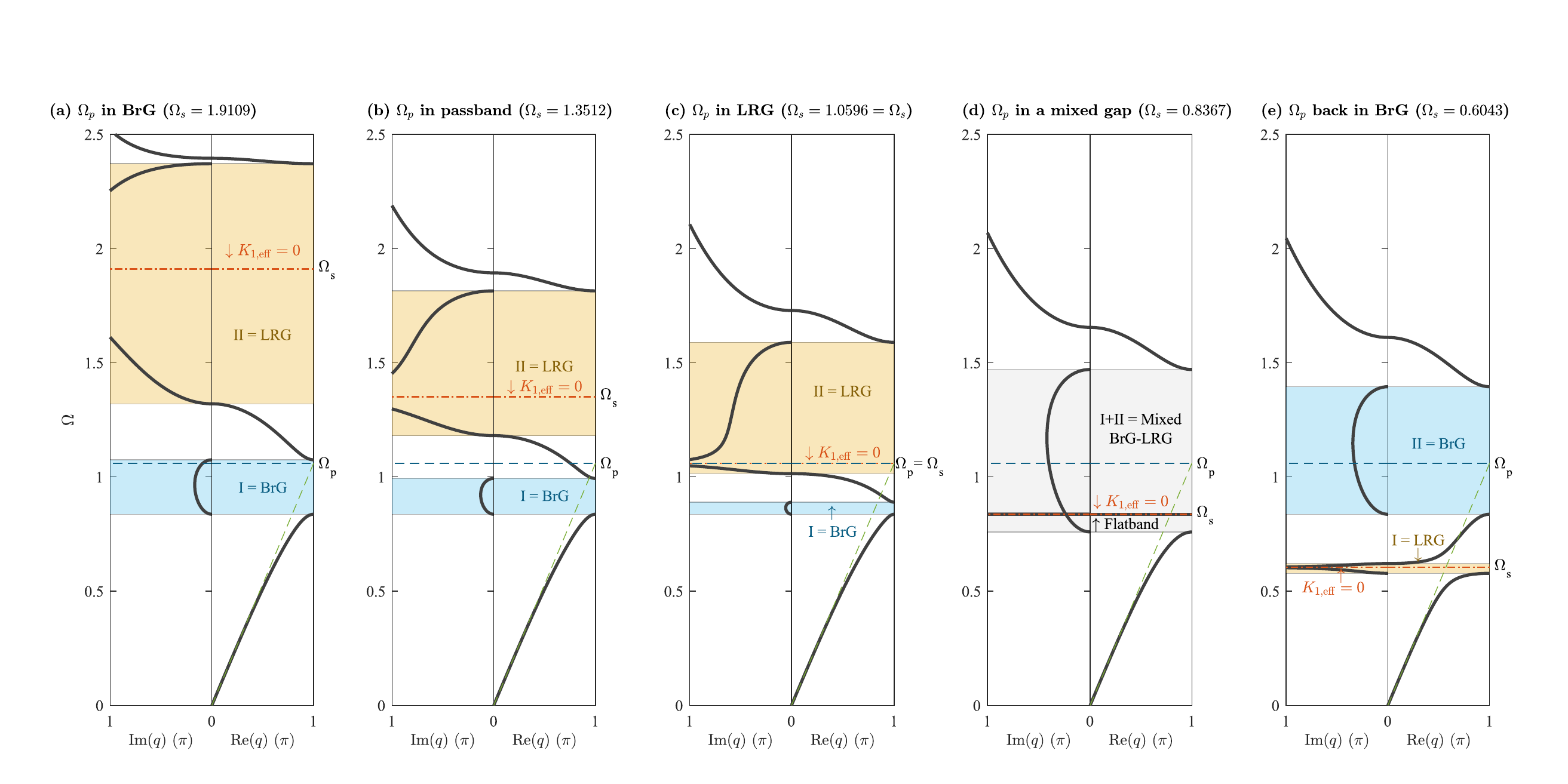}
    \caption{\kq{Bandgap switching: a migrating LRG interacts with a BrG, tracked by the characteristic frequencies $\Omega_s$ and $\Omega_p$, respectively. The slope of the green dotted line is the long-wavelength group velocity $c_{g,0}$. The LRG is always observed around $\Omega_s$, while (a) $\Omega_p$ lies in the BrG, (b) $\Omega_p$ lies in a passband, (c) $\Omega_p$ lies in the LRG, (d) $\Omega_p$ lies in the mixed BrG--LRG region, and (e) $\Omega_p$ lies in the BrG again. $\Omega_s$ is therefore a reliable diagnostic for the LRG, whereas $\Omega_p$ is unreliable for the BrG. Unit-cell parameters are $K_1=1.3$, $K_2=0.7$, $M=1$, $L/D=0.8$, $k_3=1.5$, and $m_3$ is varied through (a) $0.15$, (b) $0.3$, (c) $0.487903$, (d) $0.6043$, and (e) $1.5$, sweeping $\Omega_s$ through the indicated regimes.}}
    \label{fig:BraggLRG}
\end{figure}

\kq{One could ask whether Bragg- and local-resonance-derived features can coexist within a single gap. This raises the question of which diagnostic reliably identifies each gap type when the two mechanisms overlap. In this section we show that the local-resonance diagnostic $\Omega_s$ (the attenuation singularity at $K_{1,\mathrm{eff}}=0$) reliably tracks the LRG, whereas the Bragg diagnostic $\Omega_p$ (the periodicity frequency) does not always lie within the BrG and is therefore unreliable.

A Bragg gap forms around the periodicity frequency~\cite{kaina2013composite}
\begin{equation}
    \omega_p = \frac{\pi\, c_{g,0}}{d},
    \label{eq:omega_p}
\end{equation}
where $c_{g,0}$ is the long-wavelength group velocity of the acoustic branch and $d$ is the lattice constant. To evaluate $\omega_p$ we require $c_{g,0}$, which we obtain from the dispersion relation. The $2\times 2$ effective characteristic equation is
\begin{equation}
    M^2\omega^4 - 2M\bigl(K_{1,\mathrm{eff}}(\omega)+K_2\bigr)\,\omega^2
    + 2K_{1,\mathrm{eff}}(\omega)\,K_2\,\bigl(1-\cos qd\bigr) = 0 .
    \label{eq:char2x2}
\end{equation}
Solving for the acoustic branch,
\begin{equation}
    \omega_{\mathrm{ac}}^2
    = \frac{2M\bigl(K_{1,\mathrm{eff}}+K_2\bigr)
      - \sqrt{\,4M^2\bigl(K_{1,\mathrm{eff}}+K_2\bigr)^2
      - 8M^2 K_{1,\mathrm{eff}} K_2\,(1-\cos qd)\,}}{2M^2}.
    \label{eq:omega_acoustic}
\end{equation}
In the continuum (long-wavelength) limit $q\to 0$, the acoustic branch satisfies $\omega_{\mathrm{ac}}\to 0$, so that $K_{1,\mathrm{eff}}(\omega)\to K_{1,\mathrm{eff}}(0)=K_1$. Using $1-\cos qd \approx (qd)^2/2$ and expanding Eq.~\eqref{eq:omega_acoustic} to leading order in $q$ gives
\begin{equation}
    \omega_{\mathrm{ac}}^2 \approx
    \frac{K_1 K_2}{2M\,(K_1+K_2)}\,(qd)^2 .
    \label{eq:omega_ac_cont}
\end{equation}
The long-wavelength group velocity is therefore
\begin{equation}
    c_{g,0} = \lim_{q\to 0}\dv{\omega_{\mathrm{ac}}}{q}
    = d\,\sqrt{\frac{K_1 K_2}{2M\,(K_1+K_2)}} .
    \label{eq:cg0}
\end{equation}
Substituting Eq.~\eqref{eq:cg0} into Eq.~\eqref{eq:omega_p}, the Bragg periodicity frequency is
\begin{equation}
    \omega_p = \pi\,\sqrt{\frac{K_1 K_2}{2M\,(K_1+K_2)}},
    \label{eq:omega_p_final}
\end{equation}
and, in normalized form,
\begin{equation}
    \Omega_p = \frac{\pi}{K_1+K_2}\sqrt{\frac{K_1 K_2}{2}} .
    \label{eq:Omega_p}
\end{equation}

To test the reliability of the two diagnostics, we vary $\Omega_s$ relative to $\Omega_p$ (by tuning the resonator mass $m_3$) and track both gaps in Fig.~\ref{fig:BraggLRG}. The LRG is observed around $\Omega_s$ in every case [Fig.~\ref{fig:BraggLRG}(a)--(e)], confirming that $\Omega_s$ is a reliable diagnostic for the local-resonance gap. The behaviour of $\Omega_p$, by contrast, is not reliable: $\Omega_p$ lies within the BrG in Fig.~\ref{fig:BraggLRG}(a) and~(e), but falls inside a passband in Fig.~\ref{fig:BraggLRG}(b), inside the LRG in Fig.~\ref{fig:BraggLRG}(c), and inside the merged `mixed' BrG--LRG region in Fig.~\ref{fig:BraggLRG}(d), where the intervening band is dispersionless. The Bragg periodicity frequency $\Omega_p$ therefore does not reliably indicate the location of the BrG when the two mechanisms overlap, whereas the attenuation singularity $\Omega_s$ consistently identifies the LRG. This justifies our use of $\Omega_s$, rather than $\Omega_p$, as the primary criterion for classifying gaps and for defining the BrG-to-LRG switching reported in the main text. We further note that the LRG in Fig.~\ref{fig:BraggLRG}(e), forming entirely below $\Omega_p$, \pro{lies in the sub-Bragg regime~\cite{raghavan2013local}}.} \\

\section{Zak phase characterization}
\label{sec:zak}

In this section, we present numerical calculations of the Zak phase for both the full $3\times3$ and effective $2\times2$ representations of the intracell- and intercell-resonant stiffness dimer chains. These results confirm the inversion symmetry analysis of Sec.~\ref{sec:inversion} and demonstrate the utility of the effective framework for topological characterization.

\subsection{Zak phase calculation}
\label{sec:zak_calculation}

The Zak phase for band $b$ is computed using the discrete Wilson loop formulation:
\begin{equation}
    \theta_{\text{Zak}}^{(b)} = -\text{Im}\ln\left[\prod_{r=0}^{R-1} \langle\psi_{b}(q_{r})|\psi_{b}(q_{r+1})\rangle\right], \quad \text{with } |\psi_{b}(q_{R})\rangle \equiv |\psi_{b}(q_{0})\rangle,
    \label{eq:zak_discrete}
\end{equation}
where $|\psi_{b}(q)\rangle$ is the normalized eigenvector of the $b$th band, $q$ is the dimensionless wavenumber, $q_0 = 0$, $q_R = 2\pi$, and $R$ is the number of discretization points in the Brillouin zone. For inversion-symmetric systems, the Zak phase is quantized to either $0$ or $\pi$.

\subsection{Results for the $3\times3$ representation}
\label{sec:zak_3x3}

\begin{figure}[t!]
    \centering
    \includegraphics[width=11cm]{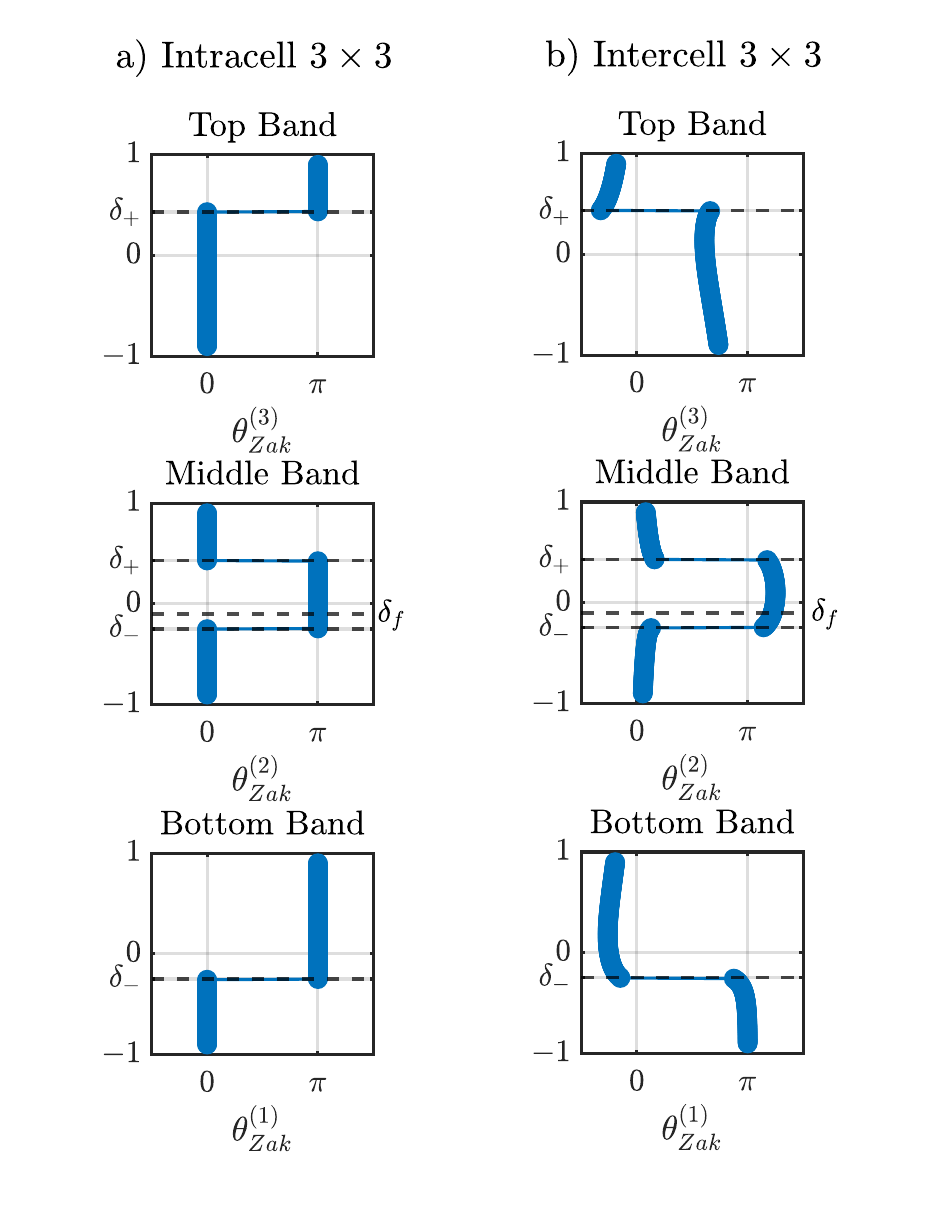}
    \caption{Band-wise Zak phase calculated for the $3\times3$ dynamical matrix of (a) the intracell-resonant stiffness dimer and (b) the intercell-resonant stiffness dimer. In the intracell system (a), the Zak phase values are quantized to either $0$ or $\pi$ due to inversion symmetry, with transitions occurring at $\delta_-$ and $\delta_+$. In the intercell system (b), the Zak phase is not quantized due to the absence of inversion symmetry in the $3\times3$ representation. Parameters: $K=1$, $M=1$, $k_3=0.43$, $m_3=0.2$, $\lambda=1$.}
    \label{fig:Zak_phase_33}
\end{figure}

Figure~\ref{fig:Zak_phase_33} shows the band-wise Zak phase calculated from the full $3\times3$ dynamical matrices as a function of $\delta$.

For the intracell-resonant configuration [Fig.~\ref{fig:Zak_phase_33}(a)], the Zak phase is quantized to $0$ or $\pi$ for all three bands, confirming the inversion symmetry established in Sec.~\ref{sec:inversion_intracell}. The Zak phase transitions occur precisely at $\delta = \delta_-$ (where bands 1 and 2 touch at DP1) and $\delta = \delta_+$ (where bands 2 and 3 touch at DP2), consistent with the band-touching analysis in Sec.~\ref{sec:band_touching}.

For the intercell-resonant configuration [Fig.~\ref{fig:Zak_phase_33}(b)], the Zak phase takes non-quantized values that vary continuously with $\delta$. This confirms the absence of inversion symmetry in the $3\times3$ representation, as established in Sec.~\ref{sec:inversion_intercell}. Consequently, the Zak phase is not a valid topological invariant for this representation.

\subsection{Results for the $2\times2$ effective representation}
\label{sec:zak_2x2}

\begin{figure}[t!]
    \centering
    \includegraphics[width=11cm]{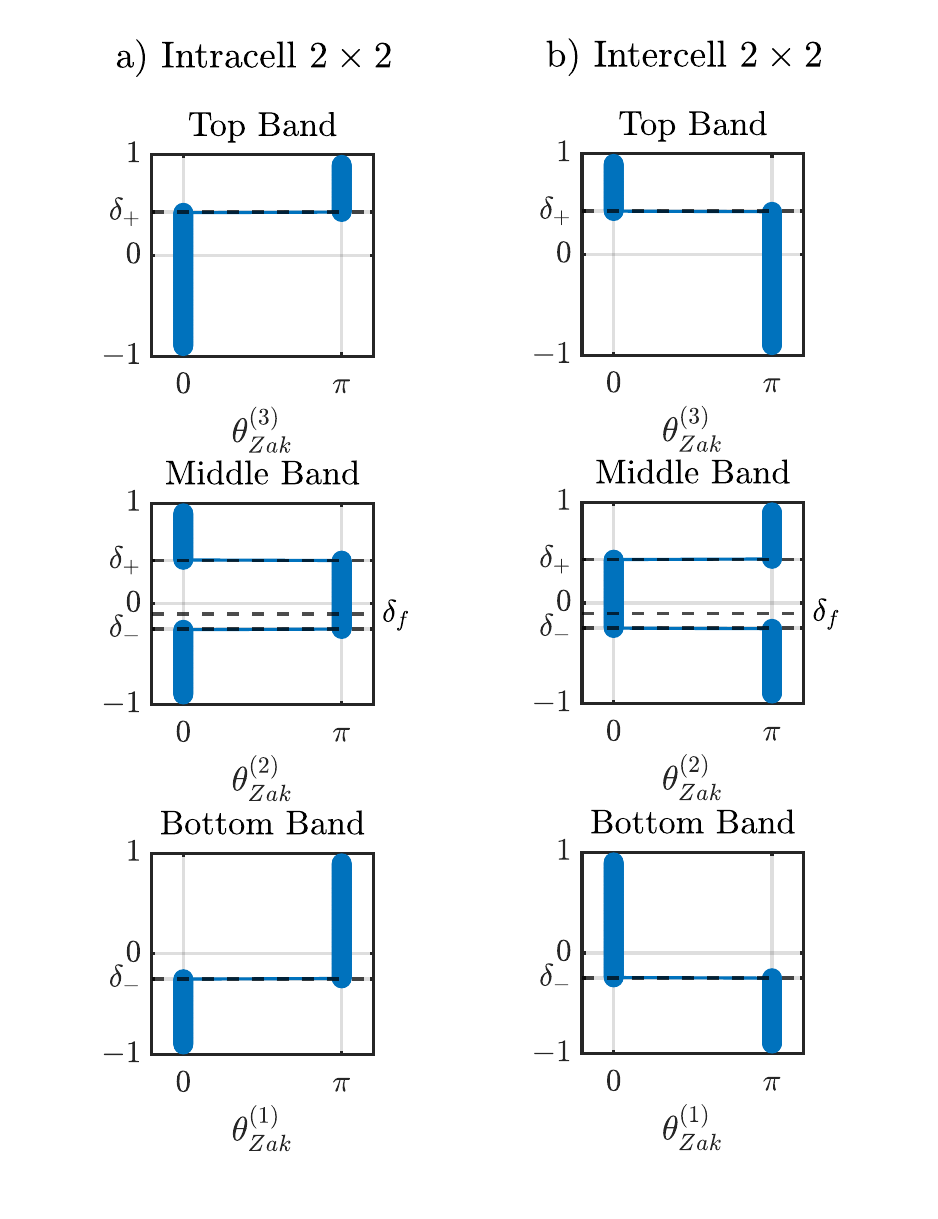}
    \caption{Band-wise Zak phase calculated for the $2\times2$ effective dynamical matrix of (a) the intracell-resonant stiffness dimer and (b) the intercell-resonant stiffness dimer. In both cases, the Zak phase values are quantized to either $0$ or $\pi$ due to inversion symmetry in the effective framework. Parameters: $K=1$, $M=1$, $k_3=0.43$, $m_3=0.2$, $\lambda=1$.}
    \label{fig:Zak_phase_2x2}
\end{figure}

Figure~\ref{fig:Zak_phase_2x2} shows the band-wise Zak phase calculated from the effective $2\times2$ dynamical matrices.

For both configurations [Figs.~\ref{fig:Zak_phase_2x2}(a) and (b)], the Zak phase is quantized to $0$ or $\pi$, confirming the inversion symmetry of the effective $2\times2$ models established in Sec.~\ref{sec:inversion_effective}. This demonstrates a key advantage of the effective framework: it enables topological characterization via the Zak phase even for systems (such as the intercell-resonant configuration) whose full $3\times3$ representation does not exhibit inversion symmetry.

\subsection{Gap topological invariant}
\label{sec:gap_invariant}

The gap topological invariant $\varsigma^{(b)}$, defined in Eq.~(22) of the main text, is computed from the sum of Zak phases of all bands below gap $b$:
\begin{equation}
    \text{sgn}[\varsigma^{(b)}] = (-1)^{b} \exp\left(i\sum_{m=1}^{b}\theta_{\text{Zak}}^{(m)}\right).
    \label{eq:gap_invariant_SM}
\end{equation}

For the intracell-resonant stiffness dimer, the results are summarized in Table~I of the main text. The key observation is that the gap-type switching at $\delta = \delta_f$ does not alter the topological invariant. This \textit{topological inheritance} mechanism enables the creation of topological LRGs: a gap that becomes non-trivial as a BrG (at $\delta > \delta_-$) retains its non-trivial character after switching to an LRG (at $\delta > \delta_f$).

\section{Frequency of Edge States}
\label{sec:finite}

In this section, we derive analytical expressions for the edge mode frequencies.

\subsection{Edge mode frequencies: Standard stiffness dimer}
\label{sec:edge_standard}

\begin{figure}[h]
    \centering
    \includegraphics[width=\textwidth]{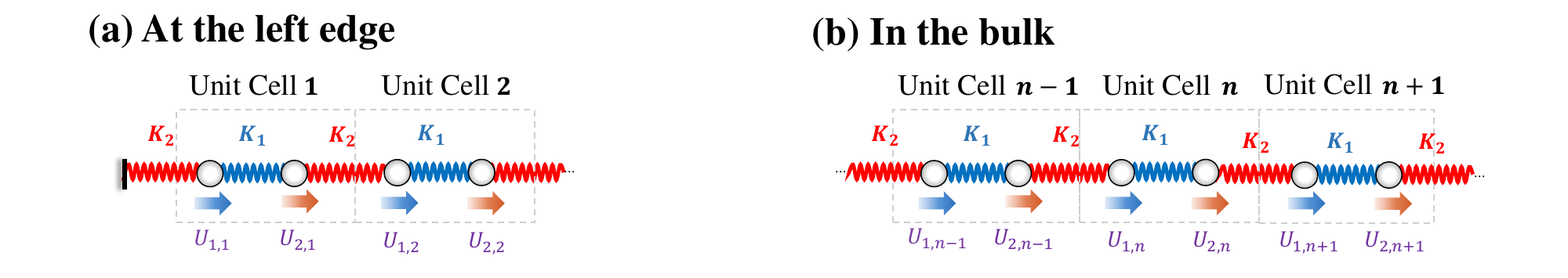}
    \caption{Finite stiffness dimer chain with fixed boundaries. (a)~Left end of the chain including the first unit cell ($j=1$). (b)~A portion of the bulk centered at unit-cell index $n$. The displacements $U_{i,j}$ are shown, where $i \in \{1,2\}$ indicates the sub-lattice index and $j$ indicates the unit-cell index.}
    \label{fig:edge_bulk}
\end{figure}

We first derive the edge mode frequency for the standard stiffness dimer chain (without local resonators), which serves as a reference. Consider a finite chain with fixed boundaries, as shown in Fig.~\ref{fig:edge_bulk}.

\subsubsection{Equations at the edge}

For the first (leftmost) unit cell, the displacements are $U_{1,1}$ and $U_{2,1}$. Assuming the edge mode decays into the bulk with a Bloch-like factor, the displacements of the second unit cell are
\begin{equation}
    U_{1,2} = U_{1,1} e^{iq}, \quad U_{2,2} = U_{2,1} e^{iq},
    \label{eq:bloch_edge}
\end{equation}
where $q$ is complex for evanescent modes.

The equations of motion for the first unit cell (with fixed left boundary, i.e., $U_{2,0} = 0$) are
\begin{align}
    M\ddot{U}_{1,1} &= -(K_1 + K_2)U_{1,1} + K_1 U_{2,1}, \label{eq:eom_edge1} \\
    M\ddot{U}_{2,1} &= -(K_1 + K_2)U_{2,1} + K_1 U_{1,1} + K_2 U_{1,2}. \label{eq:eom_edge2}
\end{align}

Assuming harmonic motion ($\ddot{U} = -\omega^2 U$) and substituting Eq.~\eqref{eq:bloch_edge} into Eq.~\eqref{eq:eom_edge2}:
\begin{align}
    M\omega^2 U_{1,1} &= (K_1 + K_2)U_{1,1} - K_1 U_{2,1}, \\
    M\omega^2 U_{2,1} &= (K_1 + K_2)U_{2,1} - K_1 U_{1,1} - K_2 U_{1,1} e^{iq}.
\end{align}

The characteristic equation is
\begin{equation}
    M^2\omega^4 - 2(K_1 + K_2)M\omega^2 + K_2^2 + K_1 K_2(2 - e^{iq}) = 0.
    \label{eq:char_edge}
\end{equation}

\subsubsection{Equations in the bulk}

For a unit cell deep in the bulk (index $n$), the displacements of neighboring cells are related by
\begin{equation}
    U_{1,n\pm1} = U_{1,n} e^{\pm iq}, \quad U_{2,n\pm1} = U_{2,n} e^{\pm iq}.
    \label{eq:bloch_bulk}
\end{equation}

The equations of motion are
\begin{align}
    M\ddot{U}_{1,n} &= -(K_1 + K_2)U_{1,n} + K_1 U_{2,n} + K_2 U_{2,n-1}, \\
    M\ddot{U}_{2,n} &= -(K_1 + K_2)U_{2,n} + K_1 U_{1,n} + K_2 U_{1,n+1}.
\end{align}

The characteristic equation is
\begin{equation}
    M^2\omega^4 - 2(K_1 + K_2)M\omega^2 + 2K_1 K_2(1 - \cos(q)) = 0.
    \label{eq:char_bulk}
\end{equation}

\subsubsection{Frequency compatibility}

For a consistent edge mode, the frequency must satisfy both the edge and bulk conditions. Equating the constant terms in Eqs.~\eqref{eq:char_edge} and \eqref{eq:char_bulk}:
\begin{equation}
    K_2^2 + K_1 K_2(2 - e^{iq}) = 2K_1 K_2(1 - \cos(q)).
    \label{eq:compatibility}
\end{equation}

Simplifying yields
\begin{equation}
    e^{-iq} = -\frac{K_2}{K_1}.
    \label{eq:wavenumber_condition}
\end{equation}

Writing $q = \text{Re}(q) + i\,\text{Im}(q)$:
\begin{equation}
    e^{-i\,\text{Re}(q)d}\, e^{\text{Im}(q)d} = -\frac{K_2}{K_1}.
    \label{eq:q_decomposition}
\end{equation}

For $K_1, K_2 > 0$, the right-hand side is negative and real, requiring
\begin{equation}
    \text{Re}(q) = \pi, \quad e^{\text{Im}(q)} = \frac{K_2}{K_1}.
    \label{eq:q_solution}
\end{equation}

For a decaying edge mode (localized at the left boundary), we need $\text{Im}(q) > 0$, which requires
\begin{equation}
    K_2 > K_1 \quad \Rightarrow \quad \delta > 0.
    \label{eq:edge_condition_left}
\end{equation}

For an edge mode localized at the right boundary, the analogous analysis gives $K_1 > K_2$, i.e., $\delta < 0$.

Substituting $q = \pi + i\ln(K_2/K_1)$ into the characteristic equation and solving:
\begin{equation}
    \omega_{\text{edge,ref}}^2 = \frac{K_1 + K_2}{M}.
    \label{eq:omega_edge_ref}
\end{equation}

With $K_1 = K(1-\delta)$ and $K_2 = K(1+\delta)$:
\begin{equation}
    \omega_{\text{edge,ref}}^2 = \frac{2K}{M}, \quad \text{or} \quad \Omega_{\text{edge,ref}}^2 = 1.
    \label{eq:Omega_edge_ref}
\end{equation}

\textbf{Key result:} The edge mode frequency in the standard stiffness dimer is independent of the dimerization parameter $\delta$.

\subsection{Edge mode frequencies: Effective stiffness dimer}
\label{sec:edge_effective}

For the effective stiffness dimer, we replace $K_1$ with the frequency-dependent effective stiffness $K_{1,\text{eff}}(\omega)$. The edge mode frequency satisfies
\begin{equation}
    \omega_{\text{edge}}^2 = \frac{K_{1,\text{eff}}(\omega_{\text{edge}}) + K_2}{M}.
    \label{eq:omega_edge_eff_condition}
\end{equation}

Substituting the expression for the effective stiffness:
\begin{equation}
    K_{1,\text{eff}}(\omega_{\text{edge}}) = K_1 + \frac{\lambda^2 k_3}{2}\frac{\omega_{\text{edge}}^2}{\omega_{\text{edge}}^2 - \omega_r^2},
    \label{eq:K1eff_edge}
\end{equation}
where $\omega_r = \sqrt{k_3/m_3}$.

Substituting into Eq.~\eqref{eq:omega_edge_eff_condition}:
\begin{equation}
    M\omega_{\text{edge}}^2 = K_1 + K_2 + \frac{\lambda^2 k_3}{2}\frac{\omega_{\text{edge}}^2}{\omega_{\text{edge}}^2 - \omega_r^2}.
    \label{eq:edge_implicit}
\end{equation}

Rearranging:
\begin{equation}
    M\omega_{\text{edge}}^2(\omega_{\text{edge}}^2 - \omega_r^2) = (K_1 + K_2)(\omega_{\text{edge}}^2 - \omega_r^2) + \frac{\lambda^2 k_3}{2}\omega_{\text{edge}}^2.
    \label{eq:edge_rearranged}
\end{equation}

This yields a quadratic equation in $\omega_{\text{edge}}^2$:
\begin{equation}
    M(\omega_{\text{edge}}^2)^2 - \left[K_1 + K_2 + M\omega_r^2 + \frac{\lambda^2 k_3}{2}\right]\omega_{\text{edge}}^2 + (K_1 + K_2)\omega_r^2 = 0.
    \label{eq:edge_quadratic}
\end{equation}

With $K_1 + K_2 = 2K$:
\begin{equation}
    M(\omega_{\text{edge}}^2)^2 - \left[2K + M\omega_r^2 + \frac{\lambda^2 k_3}{2}\right]\omega_{\text{edge}}^2 + 2K\omega_r^2 = 0.
    \label{eq:edge_quadratic_simplified}
\end{equation}

Solving:
\begin{equation}
    \omega_{\text{edge}}^2 = \frac{1}{2M}\left[\left(2K + M\omega_r^2 + \frac{\lambda^2 k_3}{2}\right) \pm \sqrt{\left(2K + M\omega_r^2 + \frac{\lambda^2 k_3}{2}\right)^2 - 8KM\omega_r^2}\right].
    \label{eq:omega_edge_solutions}
\end{equation}

This gives two edge mode frequencies:
\begin{align}
    \omega_{\text{edge,I}}^2 &= \frac{1}{2M}\left[\left(2K + M\omega_r^2 + \frac{\lambda^2 k_3}{2}\right) - \sqrt{\left(2K + M\omega_r^2 + \frac{\lambda^2 k_3}{2}\right)^2 - 8KM\omega_r^2}\right], \label{eq:omega_edge_I} \\
    \omega_{\text{edge,II}}^2 &= \frac{1}{2M}\left[\left(2K + M\omega_r^2 + \frac{\lambda^2 k_3}{2}\right) + \sqrt{\left(2K + M\omega_r^2 + \frac{\lambda^2 k_3}{2}\right)^2 - 8KM\omega_r^2}\right]. \label{eq:omega_edge_II}
\end{align}

In nondimensional form, using $\Omega^2 = \omega^2 M/(2K)$, $\Omega_r^2 = k_3 M/(2K m_3)$, and $\kappa_3 = k_3/(2K)$:
\begin{align}
    \Omega_{\text{edge,I}}^2 &= \frac{1}{2}\left[\left(1 + \Omega_r^2 + \frac{\kappa_3\lambda^2}{2}\right) - \sqrt{\left(1 + \Omega_r^2 + \frac{\kappa_3\lambda^2}{2}\right)^2 - 4\Omega_r^2}\right], \label{eq:Omega_edge_I} \\
    \Omega_{\text{edge,II}}^2 &= \frac{1}{2}\left[\left(1 + \Omega_r^2 + \frac{\kappa_3\lambda^2}{2}\right) + \sqrt{\left(1 + \Omega_r^2 + \frac{\kappa_3\lambda^2}{2}\right)^2 - 4\Omega_r^2}\right]. \label{eq:Omega_edge_II}
\end{align}
{We find that the edge mode frequencies $\Omega_{\text{edge,I}}$ in Eq.~\eqref{eq:Omega_edge_I} and $\Omega_{\text{edge,II}}$ in Eq.~\eqref{eq:Omega_edge_II}, being independent of the sweeping parameter $\delta$, are equal to $\Omega_{-}$ and $\Omega_{+}$ in Eq.~\eqref{eq:Omega_pm2}, respectively.}

\textbf{Key result:} The edge mode frequencies in the effective stiffness dimer are also independent of the dimerization parameter $\delta$. The lower edge mode (gap~I) has frequency $\Omega_{\text{edge,I}}$, and the upper edge mode (gap~II) has frequency $\Omega_{\text{edge,II}}$.

\section{Single-Particle Edge State Condition}
\label{sec:spm}

In this section, we derive the condition for the single-particle mode (SPM), where the topological edge state becomes confined to a single particle at the boundary.

\subsection{Physical mechanism}
\label{sec:spm_mechanism}

A single-particle edge state occurs when the boundary particle is effectively decoupled from the rest of the finite chain at the edge mode frequency. In the intracell-resonant configuration with fixed boundaries, this decoupling occurs when the effective intracell stiffness vanishes: $K_{1,\text{eff}}(\omega) = 0$.

At this condition, the intracell spring element (connecting the first and second particles within the unit cell) has zero effective stiffness, so the first particle oscillates independently. The natural frequency of this isolated boundary particle, connected only to the fixed wall by spring $K_2$, is
\begin{equation}
    \omega_{\text{boundary}}^2 = \frac{K_2}{M}.
    \label{eq:omega_boundary}
\end{equation}

For the SPM to exist, this boundary frequency must equal the frequency at which $K_{1,\text{eff}} = 0$, which is the attenuation singularity frequency $\omega_s$:
\begin{equation}
    \omega_s^2 = \omega_r^2 \left(1 + \frac{\lambda^2 k_3}{2K_1}\right)^{-1}.
    \label{eq:omega_s_SM}
\end{equation}

\subsection{Derivation of the SPM condition}
\label{sec:spm_derivation}

The SPM condition requires
\begin{equation}
    \omega_{\text{boundary}}^2 = \omega_s^2.
    \label{eq:spm_condition}
\end{equation}

Substituting Eqs.~\eqref{eq:omega_boundary} and \eqref{eq:omega_s_SM}:
\begin{equation}
    \frac{K_2}{M} = \frac{k_3}{m_3} \left(1 + \frac{\lambda^2 k_3}{2K_1}\right)^{-1}.
    \label{eq:spm_explicit}
\end{equation}

Rearranging:
\begin{equation}
    \frac{K_2}{M} \left(1 + \frac{\lambda^2 k_3}{2K_1}\right) = \frac{k_3}{m_3}.
    \label{eq:spm_rearranged}
\end{equation}

Substituting $K_1 = K(1-\delta)$ and $K_2 = K(1+\delta)$:
\begin{equation}
    \frac{K(1+\delta)}{M} \left(1 + \frac{\lambda^2 k_3}{2K(1-\delta)}\right) = \frac{k_3}{m_3}.
    \label{eq:spm_parametrized}
\end{equation}

Expanding:
\begin{equation}
    \frac{K(1+\delta)}{M} + \frac{\lambda^2 k_3 (1+\delta)}{2M(1-\delta)} = \frac{k_3}{m_3}.
    \label{eq:spm_expanded}
\end{equation}

Multiplying through by $2M(1-\delta)$:
\begin{equation}
    2K(1-\delta^2) + \lambda^2 k_3 (1+\delta) = \frac{2M(1-\delta)k_3}{m_3}.
    \label{eq:spm_multiplied}
\end{equation}

Expanding and collecting terms:
\begin{equation}
    2K - 2K\delta^2 + \lambda^2 k_3 + \lambda^2 k_3 \delta = \frac{2Mk_3}{m_3} - \frac{2Mk_3}{m_3}\delta.
    \label{eq:spm_expanded2}
\end{equation}

Rearranging as a quadratic in $\delta$:
\begin{equation}
    2K\delta^2 - \left(\lambda^2 k_3 + \frac{2Mk_3}{m_3}\right)\delta - \left(2K + \lambda^2 k_3 - \frac{2Mk_3}{m_3}\right) = 0.
    \label{eq:spm_quadratic}
\end{equation}

Using the quadratic formula:
\begin{equation}
    \delta = \frac{\left(\lambda^2 k_3 + \frac{2Mk_3}{m_3}\right) \pm \sqrt{\left(\lambda^2 k_3 + \frac{2Mk_3}{m_3}\right)^2 + 8K\left(2K + \lambda^2 k_3 - \frac{2Mk_3}{m_3}\right)}}{4K}.
    \label{eq:delta_s_quadratic}
\end{equation}

The discriminant simplifies to
\begin{equation}
    \left(\lambda^2 k_3 + \frac{2Mk_3}{m_3}\right)^2 + 8K\left(2K + \lambda^2 k_3 - \frac{2Mk_3}{m_3}\right) = \left(\lambda^2 k_3 + \frac{2Mk_3}{m_3} - 4K\right)^2 + 16K\lambda^2 k_3.
    \label{eq:discriminant_simplified}
\end{equation}

Therefore,
\begin{equation}
    \delta_s = \frac{1}{4K}\left[\left(\lambda^2 k_3 + \frac{2Mk_3}{m_3}\right) - \sqrt{\left(\lambda^2 k_3 + \frac{2Mk_3}{m_3} - 4K\right)^2 + 16K\lambda^2 k_3}\right],
    \label{eq:delta_s_dim}
\end{equation}
where we have chosen the minus sign to obtain a solution in the physical range $\delta \in (-1, 1)$.

In nondimensional form, using $\kappa_3 = k_3/(2K)$ and $\Omega_r^2 = k_3 M/(2K m_3)$:
\begin{equation}
    \delta_s = \left(\frac{\kappa_3\lambda^2}{2} + \Omega_r^2\right) - \sqrt{\left(\frac{\kappa_3\lambda^2}{2} + \Omega_r^2 - 1\right)^2 + 2\kappa_3\lambda^2}.
    \label{eq:delta_s_SM}
\end{equation}

\subsection{Edge mode profile transition}
\label{sec:profile_transition}

The SPM condition $\delta = \delta_s$ also marks a transition in the character of the edge mode profile:

\begin{itemize}
    \item For $\delta < \delta_s$: The edge mode decays into the bulk with alternating signs between adjacent sites (out-of-phase oscillation pattern).
    \item At $\delta = \delta_s$: The mode is confined to a single particle (IPR $= 1$\kq{, computed over the primary masses}).
    \item For $\delta > \delta_s$: The edge mode decays into the bulk with the same sign on adjacent sites (in-phase oscillation pattern).
\end{itemize}

This transition occurs because the effective stiffness $K_{1,\text{eff}}$ changes sign across the singularity:
\begin{itemize}
    \item For $\omega_{\text{edge}} < \omega_s$: $K_{1,\text{eff}} > 0$ (positive effective coupling).
    \item At $\omega_{\text{edge}} = \omega_s$: $K_{1,\text{eff}} = 0$ (decoupled).
    \item For $\omega_{\text{edge}} > \omega_s$: $K_{1,\text{eff}} < 0$ (negative effective coupling).
\end{itemize}

Since the edge mode frequency $\Omega_{\text{edge,I}}$ is fixed while the singularity frequency $\Omega_s$ varies with $\delta$, the relative position of these frequencies---and hence the sign of $K_{1,\text{eff}}$ at the edge mode frequency---changes as $\delta$ crosses $\delta_s$.

\end{document}